\documentclass[conf]{new-aiaa}
\usepackage[utf8]{inputenc}
\usepackage{tabularx}
\usepackage{multirow}
\usepackage{caption}
\usepackage{subcaption}
\usepackage{graphicx}
\usepackage{amsmath}
\usepackage[version=4]{mhchem}
\usepackage{siunitx}
\usepackage{longtable,tabularx}
\providecommand{\AnhNTAuthor}{Tuan Anh Nguyen}
\providecommand{\AnhNTAuthorEmail}{anhnt2407@konkuk.ac.kr}
\providecommand{\AnhNTAuthorAffiliation}{Konkuk Aerospace Design-Airworthiness Institute (KADA), Konkuk University, Seoul 05029, South Korea}

\providecommand{\TaehoAuthor}{Taeho Kwag}
\providecommand{\TaehoAuthorEmail}{mario0101@naver.com}
\providecommand{\TaehoAuthorAffiliation}{Korea Aerospace Industries (KAI) ltd., 78, Gongdanro 1-ro, Sanam-myeon, Sacheon-si, Gyeongsangnam-do, Korea}

\providecommand{\MinseokAuthor}{Minseok Jang}
\providecommand{\MinseokAuthorEmail}{jangminseok1996@gmail.com }
\providecommand{\MinseokAuthorAffiliation}{Korea Aerospace Industries (KAI) ltd., 78, Gongdanro 1-ro, Sanam-myeon, Sacheon-si, Gyeongsangnam-do, Korea}

\providecommand{\VinhAuthor}{Vinh Pham}
\providecommand{\VinhAuthorEmail}{quangvinhpham93@gmail.com}
\providecommand{\VinhAuthorAffiliation}{Konkuk Aerospace Design-Airworthiness Institute (KADA), Konkuk University, Seoul 05029, South Korea}

\providecommand{\NghiaAuthor}{Viet Nghia Nguyen}
\providecommand{\NghiaAuthorEmail}{nghianguyen@konkuk.ac.kr}
\providecommand{\NghiaAuthorAffiliation}{Department of Mechanical and Aerospace Engineering, Konkuk University, Seoul 05029, South Korea}

\providecommand{\JeongseokAuthor}{Jeongseok Hyun}
\providecommand{\JeongseokAuthorEmail}{ds04081@konkuk.ac.kr}

\providecommand{\JaeWooLeeAuthor}{Jae-Woo Lee}
\providecommand{\JaeWooLeeAuthorEmail}{jwlee@konkuk.ac.kr}
\providecommand{\JaeWooLeeAuthorAffiliation}{Department of Mechanical and Aerospace Engineering, Konkuk University, Seoul, South Korea}

\providecommand{\AckUAM}{This research was partially supported by Basic Science Research Program through the National Research Foundation of Korea(NRF) funded by the Ministry of Education(No. 2020R1A6A1A03046811).}

\title{AAM-VDT: Vehicle Digital Twin for Tele-Operations in Advanced Air Mobility}

\author{
\AnhNTAuthor
\footnote{Corresponding authors: \AnhNTAuthorEmail; \JaeWooLeeAuthorEmail}
\footnote{Research Professor, \AnhNTAuthorEmail},
}
\affil{\AnhNTAuthorAffiliation}

\author{
\TaehoAuthor
\footnote{Research Engineer, \TaehoAuthorEmail}
}
\affil{\TaehoAuthorAffiliation}

\author{
    \VinhAuthor
    \footnote{Postdoctoral Researcher, \VinhAuthorEmail}
}
\affil{\VinhAuthorAffiliation}

\author{
    \NghiaAuthor
    \footnote{Ph.D. Student, \NghiaAuthorEmail}
    and
    \JeongseokAuthor
    \footnote{M.Sc. Researcher, \JeongseokAuthorEmail}
}
\affil{\NghiaAuthorAffiliation}

\author{
    \MinseokAuthor
    \footnote{Research Engineer, \MinseokAuthorEmail}
}
\affil{\MinseokAuthorAffiliation}

\author{
    \JaeWooLeeAuthor
    \footnote{Professor, \JaeWooLeeAuthorEmail}    
}
\affil{\JaeWooLeeAuthorAffiliation}

\begin{document}

\maketitle

\begin{abstract}
This study advanced tele-operations in Advanced Air Mobility (AAM) through the creation of a Vehicle Digital Twin (VDT) system for eVTOL aircraft, tailored to enhance remote control safety and efficiency, especially for Beyond Visual Line of Sight (BVLOS) operations. By synergizing digital twin technology with immersive Virtual Reality (VR) interfaces, we notably elevate situational awareness and control precision for remote operators. Our VDT framework integrates immersive tele-operation with a high-fidelity aerodynamic database, essential for authentically simulating flight dynamics and control tactics. At the heart of our methodology lies an eVTOL's high-fidelity digital replica, placed within a simulated reality that accurately reflects physical laws, enabling operators to manage the aircraft via a master-slave dynamic, substantially outperforming traditional 2D interfaces. The architecture of the designed system ensures seamless interaction between the operator, the digital twin, and the actual aircraft, facilitating exact, instantaneous feedback. Experimental assessments, involving propulsion data gathering, simulation database fidelity verification, and tele-operation testing, verify the system's capability in precise control command transmission and maintaining the digital-physical eVTOL synchronization. Our findings underscore the VDT system's potential in augmenting AAM efficiency and safety, paving the way for broader digital twin application in autonomous aerial vehicles.
\end{abstract}

\section{Introduction}
\label{sec_Introduction}

Advanced Air Mobility (AAM) signifies a revolutionary leap in both urban and interurban transportation by introducing drones, air taxis, and eVTOL aircraft, promising enhanced connectivity and efficiency \cite{Licata2024}. However, it poses significant operational, maintenance, and integration challenges essential for its safe and sustainable deployment. The complexity of eVTOL aircraft and urban air mobility systems demands sophisticated remote operation technologies, leading to the inception of a digital twin-based immersive teleoperation system. This system elevates remote eVTOL operations' effectiveness, safety, and situational awareness by fusing physical and digital realms via cutting-edge simulation and virtual reality (VR) technologies, thus offering an integrated control experience for remote pilots.

The application of digital twin technology in eVTOL teleoperation ushers in a nuanced approach towards interactive and unified control ecosystems. It allows for advanced scenario simulation, predictive analytics, and informed decision-making, mitigating physical presence constraints. This research unveils an innovative teleoperation system that marries eVTOL's digital twin with a VR interface for an immersive control experience, enhancing situational awareness and enabling precision in navigating complex airspaces. At its core, the system utilizes X-In-The-Loop (XITL) interfaces—Software-In-The-Loop (SITL) and Hardware-In-The-Loop (HITL)—ensuring accurate flight dynamics and operational validation by harmonizing physical and digital layers \cite{Ywet2024}. Integral to this framework is a meticulously developed Aerodynamic Database (AeroDB), achieved by merging varied analyses to simulate authentic flight conditions efficiently, thus redefining the paradigms of remote eVTOL operational capabilities, safety, efficiency, and scalability. This advancement in AAM teleoperation, leveraging immersive VR and a robust AeroDB, marks a significant milestone, paving the way for future digital twin technology integration in unmanned aerial systems.

\begin{figure*}[!tbp]
    \centering
    \begin{subfigure}[b]{\linewidth}
        \centering
        \includegraphics[width=0.5\textwidth]{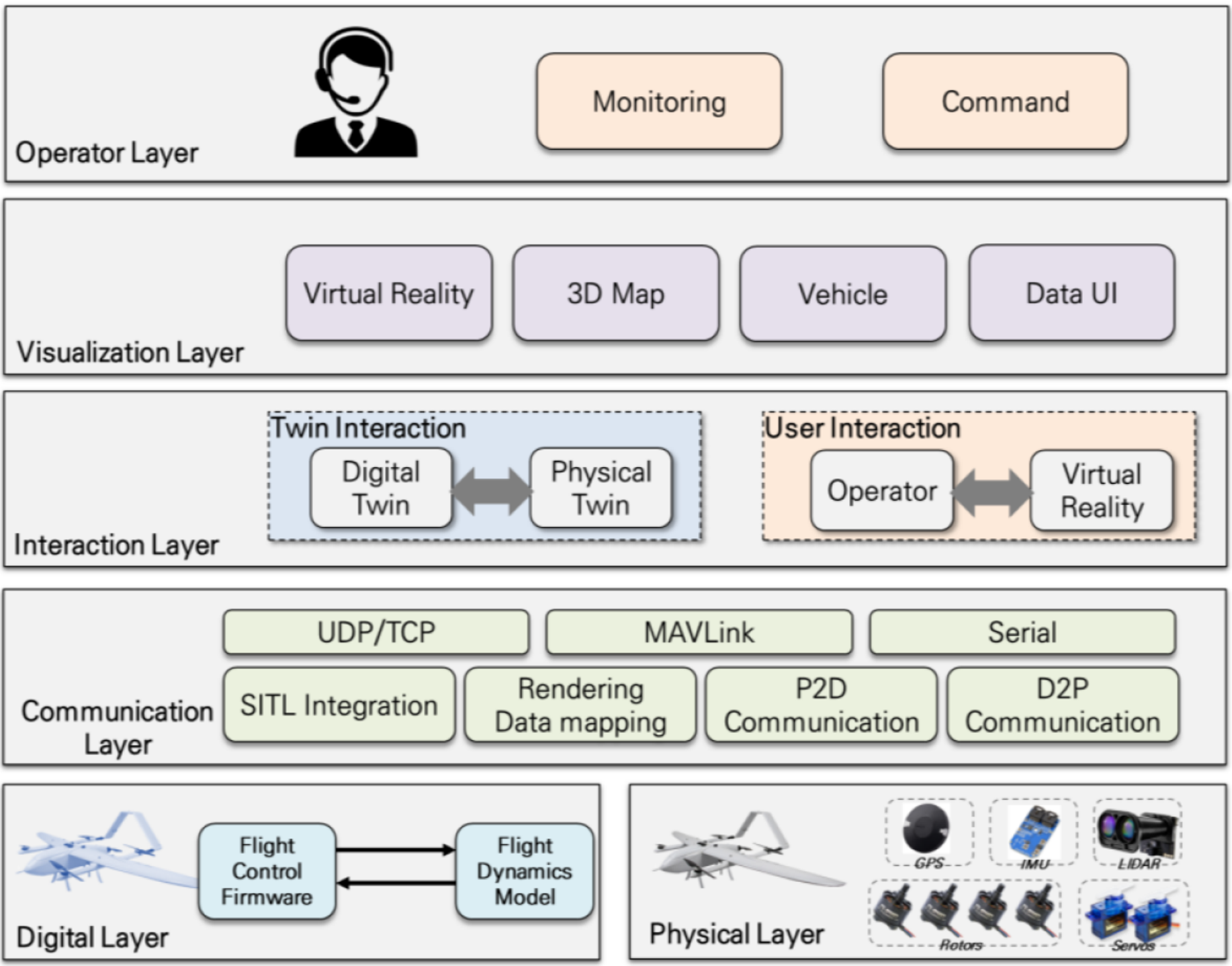}
        \caption{Functional Layers}
        \label{fig_VDT_D2P_Layer_Diagram_Funtionalities}
    \end{subfigure}
    \hfill%
    \begin{subfigure}[b]{\linewidth}
        \centering
        \includegraphics[width=0.75\textwidth]{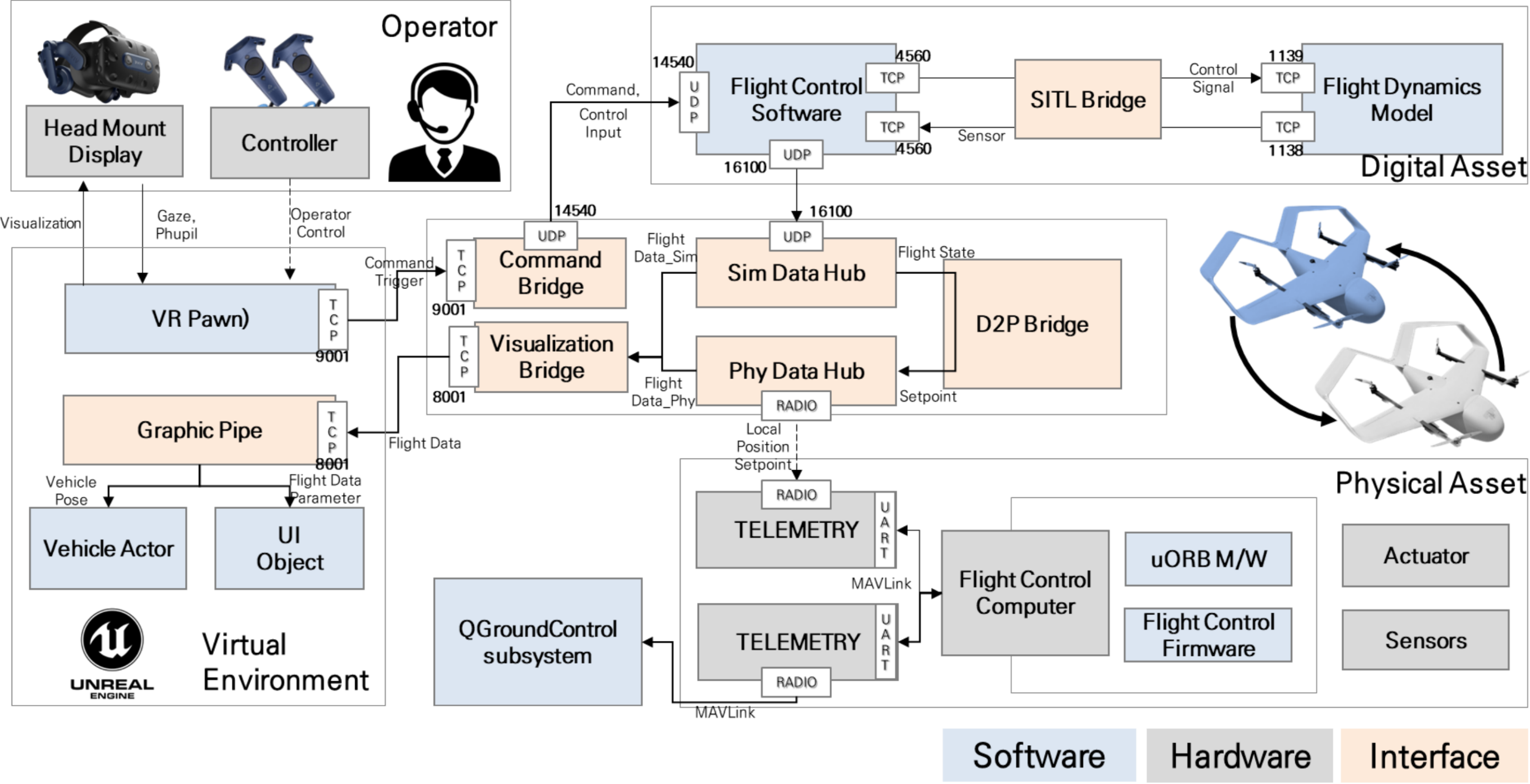}
        \caption{VDT System Architecture}
        \label{fig_VDT_D2P_Architecture}
    \end{subfigure}
    \caption{Vehicle Digital Twin Framework}
    \label{fig_VDT_Framework}
    \vspace{-10pt}
\end{figure*}

\subsection{Framework Proposal}

We propose an architecture for an immersive teleoperation system using digital twin technology, illustrated in Figure \ref{fig_VDT_Framework}. We defined crucial functions and components, orchestrating data flow across modules to formulate the architecture, with a layer diagram (Figure \ref{fig_VDT_D2P_Layer_Diagram_Funtionalities}) clarifying the concept of controlling a physical aircraft via its digital twin. The Operator Layer enables immersive monitoring and command input through a VR interface, facilitating observation from varied perspectives. The Visualization Layer, utilizing the Unreal engine, crafts a virtual environment that includes a 3D map and GUI for data visualization, enriched with VR for immersive interaction. The Interaction Layer supports bidirectional communication between digital and physical realms, enhancing operator immersion. Communication protocols within the Communication/Interface Layer ensure data exchange, employing MAVLink for real-time updates and TCP for graphical data. The architecture is divided into digital and physical layers, incorporating the eVTOL and its systems, and digital flight control simulations, respectively. This division allows for precise UAS operation simulations. The VDT system architecture, detailed in Figure \ref{fig_VDT_D2P_Architecture}, segregates into the digital layer for operator interaction and the physical eVTOL environment. Operators use a VR interface to engage with the digital twin in an Unreal-rendered virtual environment, facilitated by HMD displays for a comprehensive experience. This integration employs PX4 flight control software and KFDS for flight dynamics simulation via an SITL interface, ensuring accurate feedback and control. The D2P bridge translates simulated data into commands for the actual eVTOL, enabling seamless digital-physical aircraft interaction.

\subsection{AAM Vehicle Configuration and Modeling}
\label{sec_AAM_Vehicle_Configuration_and_Modeling}

\begin{figure*}[htbp]
    \footnotesize
    \centering
    \begin{subfigure}[b]{0.455\textwidth}
        \centering
        \includegraphics[width=1.0\textwidth]{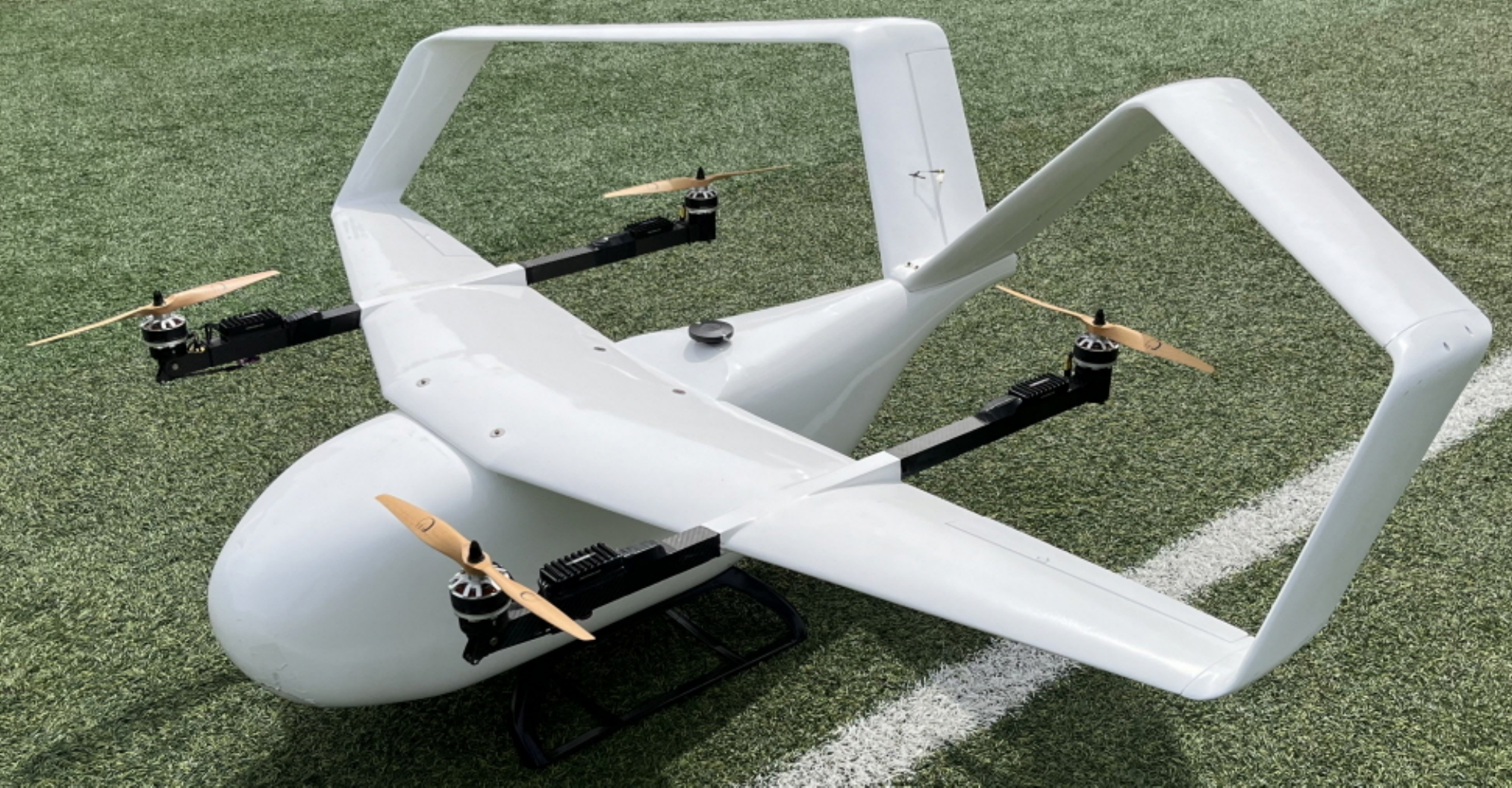}
        \caption{}
        \label{fig_VDT_Vehicle_Real}
    \end{subfigure}
    \hfill 
    \begin{subfigure}[b]{0.455\textwidth}
        \centering
        \includegraphics[width=1.0\textwidth]{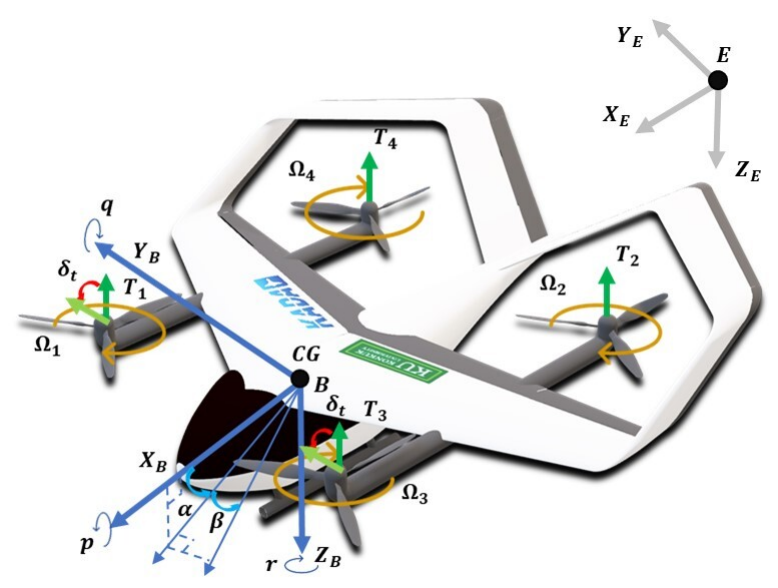}
        \caption{}
        \label{fig_VDT_Vehicle_Coordination_Full}
    \end{subfigure}
    \hfill 
    \begin{subfigure}[b]{0.33\textwidth}
        \centering
        \includegraphics[width=1.0\textwidth]{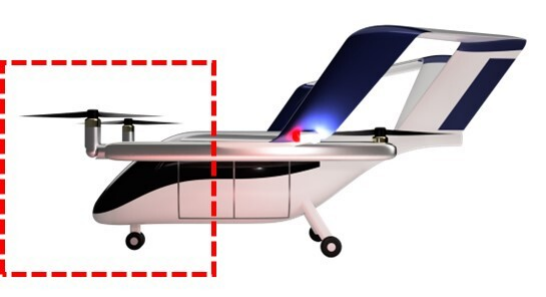}
        \caption{}
        \label{fig_VDT_Vehicle_VTOL_Mode}
    \end{subfigure}%
    \hfill 
    \begin{subfigure}[b]{0.33\textwidth}
        \centering
        \includegraphics[width=1.0\textwidth]{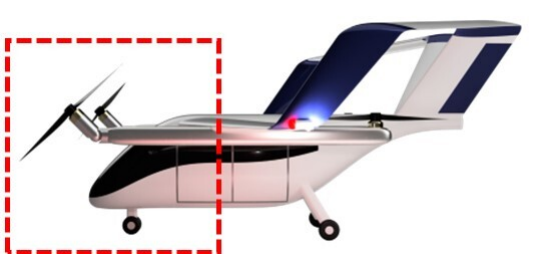}
        \caption{}
        \label{fig_VDT_Vehicle_Transition_Mode}
    \end{subfigure}%
    \hfill 
    \begin{subfigure}[b]{0.33\textwidth}
        \centering
        \includegraphics[width=1.0\textwidth]{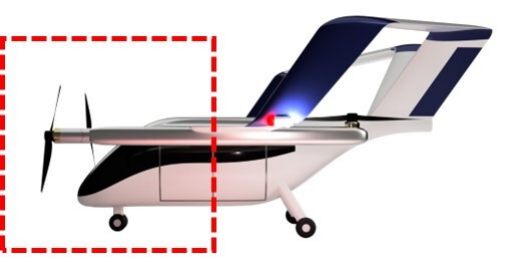}
        \caption{}
        \label{fig_VDT_Vehicle_CTOL_Mode}
    \end{subfigure}
    \caption{
        KP-2 Scaled Model’s Configuration and Flight Modes: 
        \footnotesize 
        (a) Real-world scaled model of KP-2 ($\sim$ \texttt{11.828 kg})
        (b) Body Frame and Coordinates
        (c) VTOL mode $(\delta_{t} = {90}^{\circ})$
        (d) Tilting mode $({0}^{\circ} \leq  \delta_{t} \leq {90}^{\circ})$
        (e) Cruising mode $(\delta_{t} = {0}^{\circ})$
    }
    \label{fig_VDT_Vehicle_Configuration_and_Flight_Mode}
    \vspace{-10pt}
\end{figure*}

The KP-2 AAM vehicle, depicted in Figure \ref{fig_VDT_Vehicle_Configuration_and_Flight_Mode}, adopts a boxed-wing architecture, enhancing lift efficiency beyond conventional aircraft norms, and incorporates ailerons and ruddervators for agility. This configuration, notable for its extensive wingtip area, is engineered to minimize lateral forces using four vertically mounted thrust rotors. Visual and technical details of the KP-2 are presented in Figures \ref{fig_VDT_Vehicle_Real}, \ref{fig_VDT_Vehicle_Coordination_Full}, and Table \ref{tbl_KP2_spec}. It operates across multicopter, transition, and fixed-wing modes, as depicted in Figures \ref{fig_VDT_Vehicle_VTOL_Mode} through \ref{fig_VDT_Vehicle_CTOL_Mode}, with mode transitions regulated by the front rotor's tilt from vertical (\texttt{90}$^{\circ}$) to horizontal (\texttt{0}$^{\circ}$) orientations. This design facilitates lift in vertical takeoff, forward propulsion in transition, and aerodynamic stability in horizontal flight, employing distinct rotor functions and control surfaces. Flight dynamics and the essential cross-inertial component ${I_{x z}}$ are articulated in Equation \ref{eq_VDT_Vehicle_Dynamics_Final_Model} \cite{Jang2023}, highlighting the vehicle's complex maneuvering capabilities across its operational spectrum.

\begin{table}[!htbp]
    \footnotesize
    \centering
    \caption{Specification of KP-2}
    \label{tbl_KP2_spec}
    \begin{tabular}{ccc}
        \hline
        \textbf{Parameters} & \textbf{Value} & \textbf{Unit} \\
        \hline
        MTOW & 11.828 & kg\\
        \hline
        Wingspan & 2.0 & m\\
        \hline
        Wing Area & 0.8544 & \text{m}$^2$\\
        \hline
        Fuselage Length & 1.4 & m\\
        \hline
        Aspect ratio & 7.19 \\
        \hline
        Mean aerodynamic chord & 0.2995 \\
        \hline
        Cruise Speed & 25 & m/s\\
        \hline
        Moments of Inertia $I_{xx}$ & 0.7816 & kg$\cdot$\text{m}$^2$ \\
        \hline
        Moments of Inertia $I_{yy}$ & 2.073 & kg$\cdot$\text{m}$^2$\\
        \hline
        Moments of Inertia $I_{zz}$ & 1.423 & kg$\cdot$\text{m}$^2$\\
        \hline
        Moments of Inertia $I_{xy}$ & 0.0 & kg$\cdot$\text{m}$^2$\\
        \hline
        Moments of Inertia $I_{xz}$ & $-$0.1564 & kg$\cdot$\text{m}$^2$\\
        \hline
        Moments of Inertia $I_{yz}$ & 0.0 & kg$\cdot$\text{m}$^2$\\
        \hline
    \end{tabular}
\end{table}

\begin{equation}
    \label{eq_VDT_Vehicle_Dynamics_Final_Model}
    \begin{cases}
        \dot{u}&=r v-q w-g \sin \theta+\frac{F_{x}}{m}\\
        \dot{v}&=p w-u r+\mathrm{g} \sin \phi \cos \theta+\frac{F_{y}}{m}\\
        \dot{w}&=q u+p v+g \cos \phi \cos \theta+\frac{F_{z}}{m}\\
        \dot{p}&=\frac{I_{z z} I_{x x}}{I_{z z} I_{x x}-I_{x z}{ }^{2}} (\frac{I_{z z}-I_{y y}}{I_{x x}} q r+\frac{I_{z z}-I_{y y}}{I_{x x}} \cdot \frac{I_{x z}}{I_{x x}} p q-\frac{I_{x z}{ }^{2}}{I_{z z} I_{x x}} q r+\\
        &\frac{I_{x z}}{I_{z z} I_{x x}} N+\frac{I_{x z}}{I_{x x}} p q+\frac{L}{I_{x x}})\\
        \dot{q}&=\frac{I_{z z}-I_{x x}}{I_{y y}} p r+\left(r^{2}-p^{2}\right) \frac{I_{x z}}{I_{y y}}+\frac{M}{I_{y y}}\\
        \dot{r}&=(\frac{I_{z z} I_{x x}}{I_{z z} I_{x x}-I_{x z}{ }^{2}}(\frac{I_{x x}-I_{y y}}{I_{z z}} p q+\frac{I_{y y}-I_{z z}}{I_{x x}} \cdot \frac{I_{x z}}{I_{z z}} q r+\frac{I_{x z}^{2}}{I_{z z} I_{x x}} p q+\\
        &\frac{I_{x z}}{I_{z z} I_{x x}} L-\frac{I_{x z}}{I_{z z}} q r+\frac{N}{I_{z z}})
    \end{cases}
\end{equation}

\subsection{Immersive Tele-Operation}
\label{sec_teleoperation}

This section elucidates the execution of Digital-to-Physical (D2P) command conveyance within an immersive teleoperation system, utilizing the digital simulation layer to interact with an eVTOL aircraft's physical layer. D2P control uniquely transmits control inputs and state data from the digital simulation directly to the physical aircraft, influencing its maneuvers based on the simulated state. This process extracts and transmits the simulated state information to the aircraft, informing its subsequent actions.

The system integrates flight control aspects—position, velocity, and attitude—with added logic for assimilating received state data into the flight computer's controls. Control inputs or waypoints generate a Position Setpoint, subsequently refined through controllers to derive Velocity and Acceleration Setpoints, transformed into PWM signals for actuators using simulation-derived Setpoints for real-time control, with a focus on speed value integration.

\begin{figure}[!htbp]
    \centering
    \begin{subfigure}{\linewidth}
        \centering
        \includegraphics[width=0.75\textwidth]{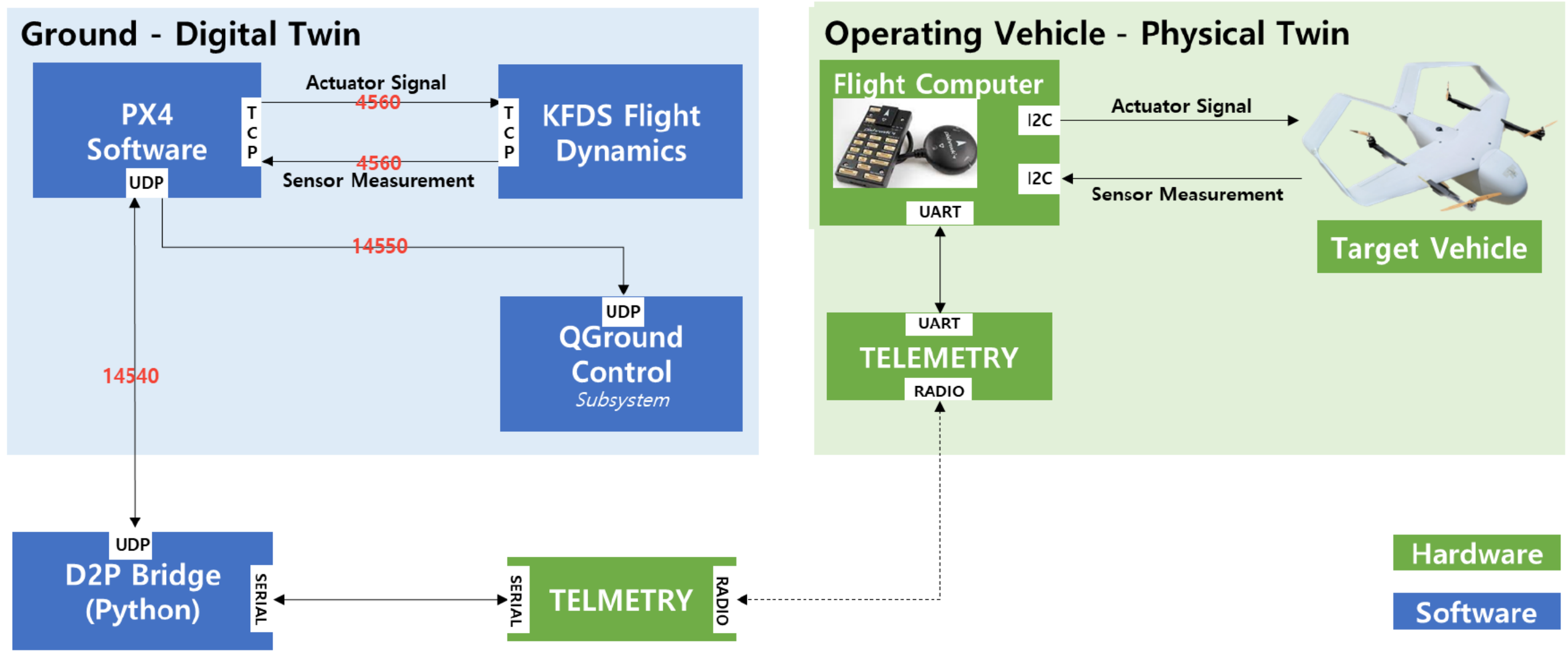}
        \caption{D2P System Configuration Diagram}
        \label{fig_VDT_D2P_System_Configuration_Diagram}
    \end{subfigure}%
    \hfill%
    \begin{subfigure}{\linewidth}
        \centering
        \includegraphics[width=0.5\textwidth]{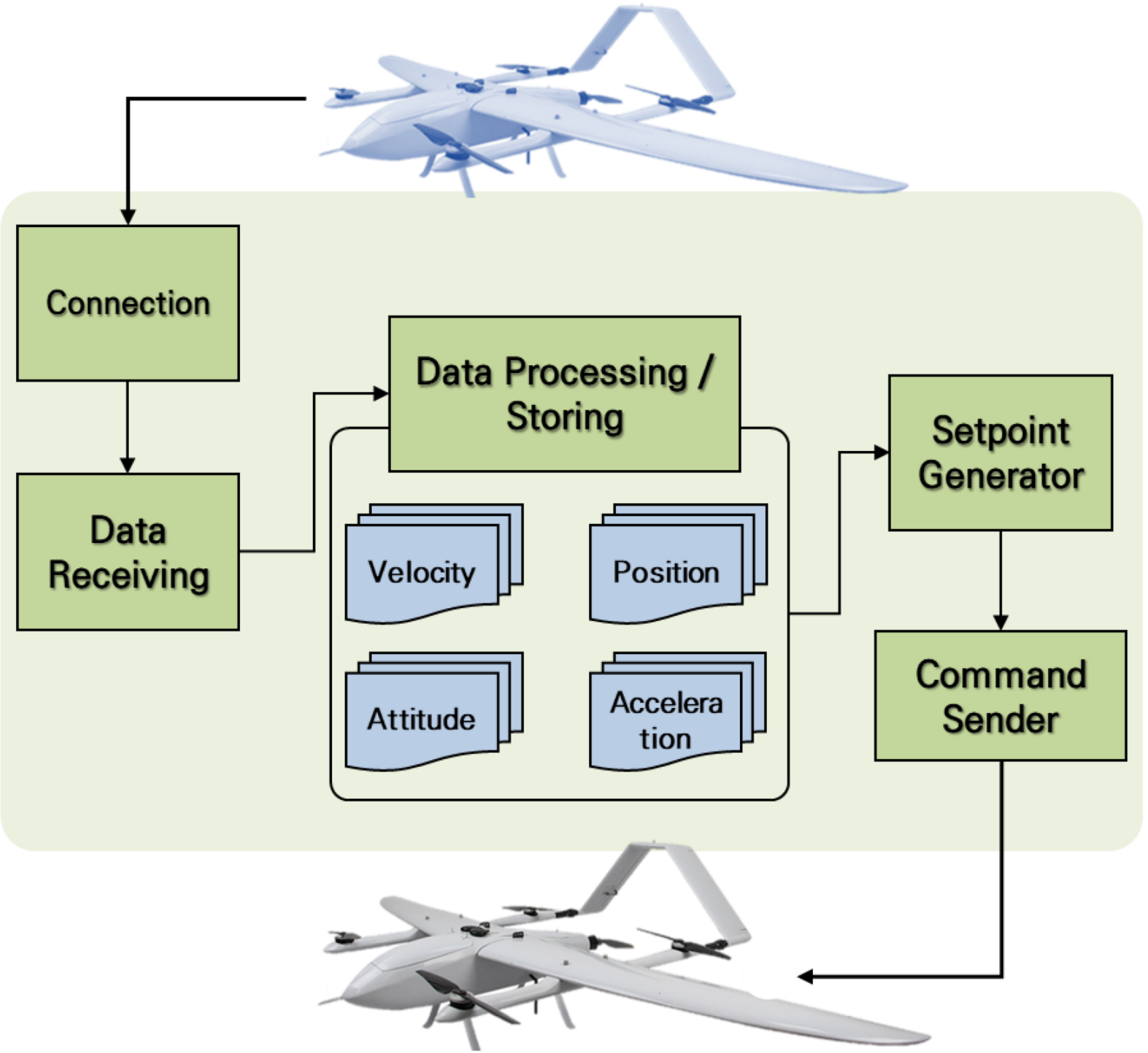}
        \caption{Sequence diagram of data processing in the D2P bridge}
        \label{fig_VDT_D2P_Sequence_Data_Processing_D2P_Bridge}
    \end{subfigure}%
    \caption{Digital to Physical System Integration}
    \label{fig_VDT_D2P_System}
    \vspace{-10pt}
\end{figure}

D2P in-flight communication employs PX4's Offboard mode to apply Setpoints using digital simulation state data, facilitated by a ground server for digital twin stability. The D2P bridge structure, showcased in Figure \ref{fig_VDT_D2P_Sequence_Data_Processing_D2P_Bridge}, leverages MAVLink for data exchange, specifically using $SET\_POSITION\_TARGET\_LOCAL\_NED$ for position and velocity, ensuring Offboard mode's reliability through a $30 \mathrm{~Hz}$ data streaming rate, affirming D2P communication's efficacy within the teleoperation paradigm.

\subsection{High-Fidelity Aerodynamic Database Generation}
\label{sec_datafusion}

This research develops an Aerodynamic Database (AeroDB) for the KP-2 aircraft, enabling detailed flight simulations for varied scenarios. Starting with specific conditions, it assesses forces and moments through table lookups to predict trajectory, integrating databases on aerodynamic coefficients, geometry, mass, and propulsion. AeroDB, crucial for its comprehensive aerodynamic data, offers precise three-dimensional force and moment information across different flight conditions and configurations, utilizing a component accumulation method for force and moment estimation, vital for accurate flight simulation.

\begin{equation}
    \begin{aligned}
        C_i=C_{i 0}(\alpha, \beta, M) & +\Delta C_{i q}(\alpha, M, q) \cdot \frac{c q}{2 U_{\infty}} \\
        & +\Delta C_{i \delta}(\alpha, M, \delta) \cdot \delta
    \end{aligned}  
\end{equation}

when $i=L, D, m$

\begin{equation}
    \begin{aligned}
        C_i=&C_{i 0}(\alpha, \beta, M) +\Delta C_{i p}(\alpha, M, p) \cdot \frac{b p}{2 U_{\infty}}+ 
        \\
        & \Delta C_{i r}(\alpha, M, r) \cdot \frac{c r}{2 U_{\infty}} +\Delta C_{i \delta}(\alpha, M, \delta) \cdot \delta
    \end{aligned}    
\end{equation}

For $i=Y_w, l, n$, $C_{i 0}$ represent primary clean configuration coefficients, while $\Delta C_{i q}, \Delta C_{i p}, \Delta C_{i r}$, and $\Delta C_{i \delta}$ denote dynamic and control coefficients for surfaces like elevators, ailerons, and rudders, respectively. Each coefficient type is mapped to lookup tables, varying from 9 to 50 based on configurations and effects, underscoring the need for efficient numerical methods to construct comprehensive databases affordably, as illustrated in Table \ref{tbl_AeroDB_format}.

\begin{table}[!htbp]
    \centering    
    \caption{Format of the AeroDB}
    \label{tbl_AeroDB_format}
    \begin{tabularx}{\textwidth}{*{15}{>{\centering\arraybackslash}X}} 
        \hline 
        $\boldsymbol{\alpha}$ & $\boldsymbol{\beta}$ & $\boldsymbol{\delta}_{\text{rud}}$ & $\boldsymbol{\delta}_{\text{rud}}$ & $\boldsymbol{\delta}_{\text{rud}}$ & $\cdots$ & $\boldsymbol{p}$ & $\boldsymbol{q}$ & $\boldsymbol{r}$ & $\boldsymbol{C}_{\boldsymbol{L}}$ & $\boldsymbol{C}_{\boldsymbol{D}}$ & $\boldsymbol{C}_{\boldsymbol{m}}$ & $\boldsymbol{C}_{\boldsymbol{Y}}$ & $\boldsymbol{C}_{\boldsymbol{l}}$ & $\boldsymbol{C}_{\boldsymbol{n}}$ \\
        \hline
        x & x & x & - & - & - & - & - & - & x & x & x & x & x & x \\
        \hline
        x & x & - & x & - & - & - & - & - & x & x & x & x & x & x \\
        \hline
        x & x & - & - & x & - & - & - & - & x & x & x & x & x & x \\
        \hline
        x & x & - & - & - & x & - & - & - & x & x & x & x & x & x \\
        \hline
        x & x & - & - & - & - & x & - & - & x & x & x & x & x & x \\
        \hline
        x & x & - & - & - & - & - & x & - & x & x & x & x & x & x \\
        \hline
        x & x & - & - & - & - & - & - & x & x & x & x & x & x & x \\
        \hline
    \end{tabularx}
    {\raggedleft \textit{“x” indicates a column vector of non-zero elements}\par}
\end{table}

\paragraph{AeroDB Generation Framework} Aerodynamic coefficients are derived from configuration data via computational methods (empirical analysis, panel methods, RANS CFD) and experimental approaches (wind tunnel tests, flight experiments), contributing to AeroDB's development. Low-fidelity analyses utilize simplified methods for swift outcomes, whereas high-fidelity analyses rely on finite difference equations, requiring extensive computational efforts. Data fusion post multi-fidelity analysis merges datasets, optimizing the database's construction by balancing computational efficiency with precision.

\paragraph{Proposed Data Fusion Method} This research employs the Extended Hierarchical Kriging (EHK) method for data fusion, integrating one high-fidelity (HF) dataset with multiple low-fidelity (LF) datasets \cite{PHAM2023, Pham2023_02}. The EHK formula approximates the HF model as:
\begin{equation}
    \hat{f}_{HF}(x)=\boldsymbol{\rho}^T \hat{\boldsymbol{f}}_{LF}(x)+Z_d(x),    
\end{equation}
where $\hat{\boldsymbol{f}}_{LF}(x)$ denotes the LF model vectors, scaled by unknown constants $\boldsymbol{\rho}$, and $Z_d(x)$ signifies the discrepancy model with mean zero and variance $\sigma_d^2$. The EHK model's prediction at any new point $x$ is:
\begin{equation}
    \hat{f}_{HF}(x)=\boldsymbol{\rho}^T \hat{\boldsymbol{f}}_{LF}(x)+\boldsymbol{r}(x)^T \boldsymbol{R}^{-1}(\boldsymbol{y}_{HF}-\boldsymbol{\rho}^T \widehat{\boldsymbol{F}}_{LF}),
\end{equation}
highlighting the correlation between estimated and actual HF responses, with $\widehat{\boldsymbol{F}}_{LF}$ being the LF response matrix at HF sample points, $\boldsymbol{R}$ the HF samples' correlation matrix, and $\boldsymbol{r}(x)$ the correlation vector to HF training samples, determined by the spatial correlation function $\phi$ based on Euclidean distances. The well-known Gaussian exponential function in second order is employed,
\begin{equation}
    \begin{gathered}
        \boldsymbol{R}\left(x_{H F}, x_{H F}^{\prime}\right)=\left(\phi\left(x_{H F}, x_{H F}^{\prime}\right)\right)_{i, j} \in \mathbb{R}^{n_{H F} \times n_{H F}} \\
        \phi\left(x, x^{\prime}\right)=\prod_{k=1}^m \exp \left(-\theta_k\left|x^{(k)}-x^{\prime(k)}\right|^2\right)
    \end{gathered}
\end{equation}
Where $\boldsymbol{\theta}=\left[\theta_1, \theta_2, \ldots, \theta_m\right] \in \mathbb{R}^m$ is a vector of unknown hyper-parameters in the Gaussian exponential function.
To estimate the hyper-parameters $\boldsymbol{\rho}, \boldsymbol{\theta}$, and $\sigma_d^2$, the maximum likelihood estimation method is employed to maximize the likelihood function given by:
\begin{equation}
    \begin{aligned}
        & \ln \{L(\theta)\}=-n_{H F} \ln \left(\sigma_d^2\right)-\ln |\boldsymbol{R}| \\
        &-\frac{\left(\boldsymbol{y}_{H F}-\boldsymbol{\rho}^{\mathrm{T}} \widehat{\boldsymbol{F}}_{L F}\right)^{\mathrm{T}} \boldsymbol{R}^{-1}\left(\boldsymbol{y}_{H F}-\boldsymbol{\rho}^{\mathrm{T}} \widehat{\boldsymbol{F}}_{L F}\right)^{\mathrm{T}}}{\sigma_d^2}
    \end{aligned}    
\end{equation}

The unknown scaling factors and process variances are analytically calculated as follows:

\begin{equation}
    \begin{gathered}
        \rho(\theta)=\left(\hat{F}_{L F}{ }^{\mathrm{T}} R^{-1} \hat{F}_{L F}\right)^{-1} \hat{F}_{L F}{ }^{\mathrm{T}} R^{-1} y_{H F} \\
        \sigma_d^2(\theta)=\frac{1}{n_{H F}}\left(y_{H F}-\rho^{\mathrm{T}} \hat{F}_{L F}\right)^{\mathrm{T}} R^{-1}\left(y_{H F}-\rho^{\mathrm{T}} \hat{F}_{L F}\right)
    \end{gathered}   
\end{equation}

\begin{figure*}[!htbp]
    \centering
    \begin{subfigure}{0.455\linewidth}
        \centering
        \includegraphics[width=0.88\linewidth]{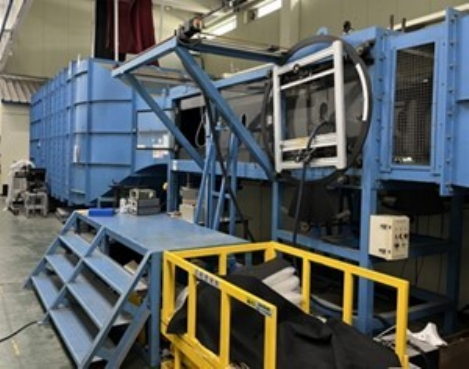}
        \caption{Konkuk University's Wind Tunnel}
        \label{fig_VDT_Wind_Tunnel_Test_Equipment}
    \end{subfigure}%
    \hfill%
    \begin{subfigure}{0.455\linewidth}
        \centering
        \includegraphics[width=0.85\linewidth]{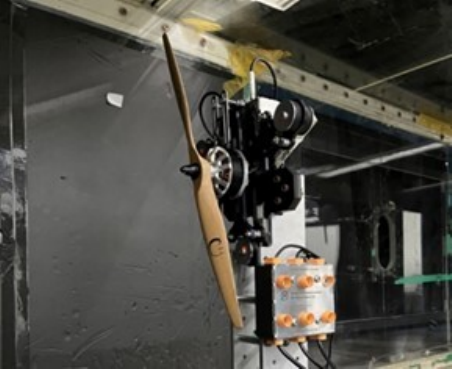}
        \caption{Test-bed for Propulsion Module}
        \label{fig_VDT_Wind_Tunnel_Test_Propeller_Setting}
    \end{subfigure}%
    \hfill%
    \begin{subfigure}{0.455\linewidth}
        \centering
        \includegraphics[width=1.0\linewidth]{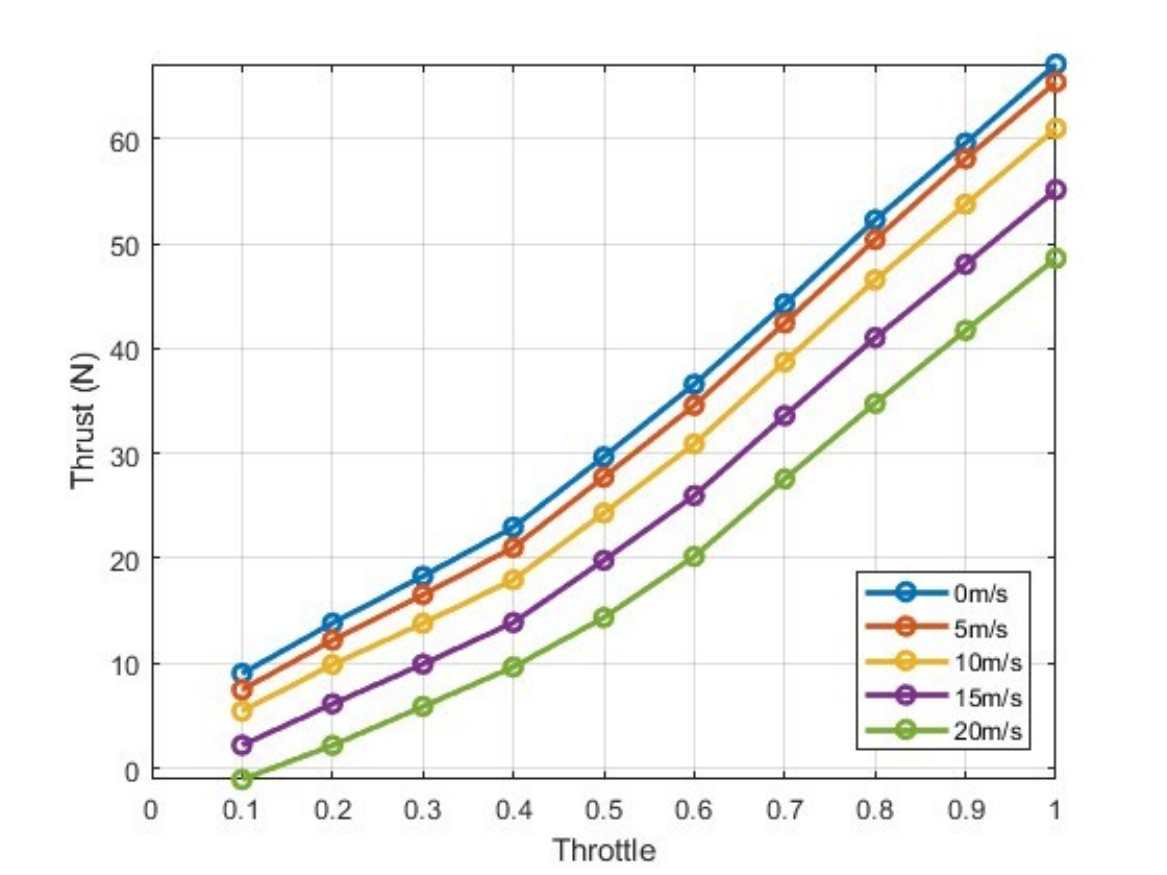}
        \caption{Thrust wrt. Inflow Speed}
        \label{fig_VDT_Wind_Tunnel_Test_Thrust_Inflowspeed_Plot}
    \end{subfigure}%
    \hfill%
    \begin{subfigure}{0.455\linewidth}
        \centering
        \includegraphics[width=1.0\linewidth]{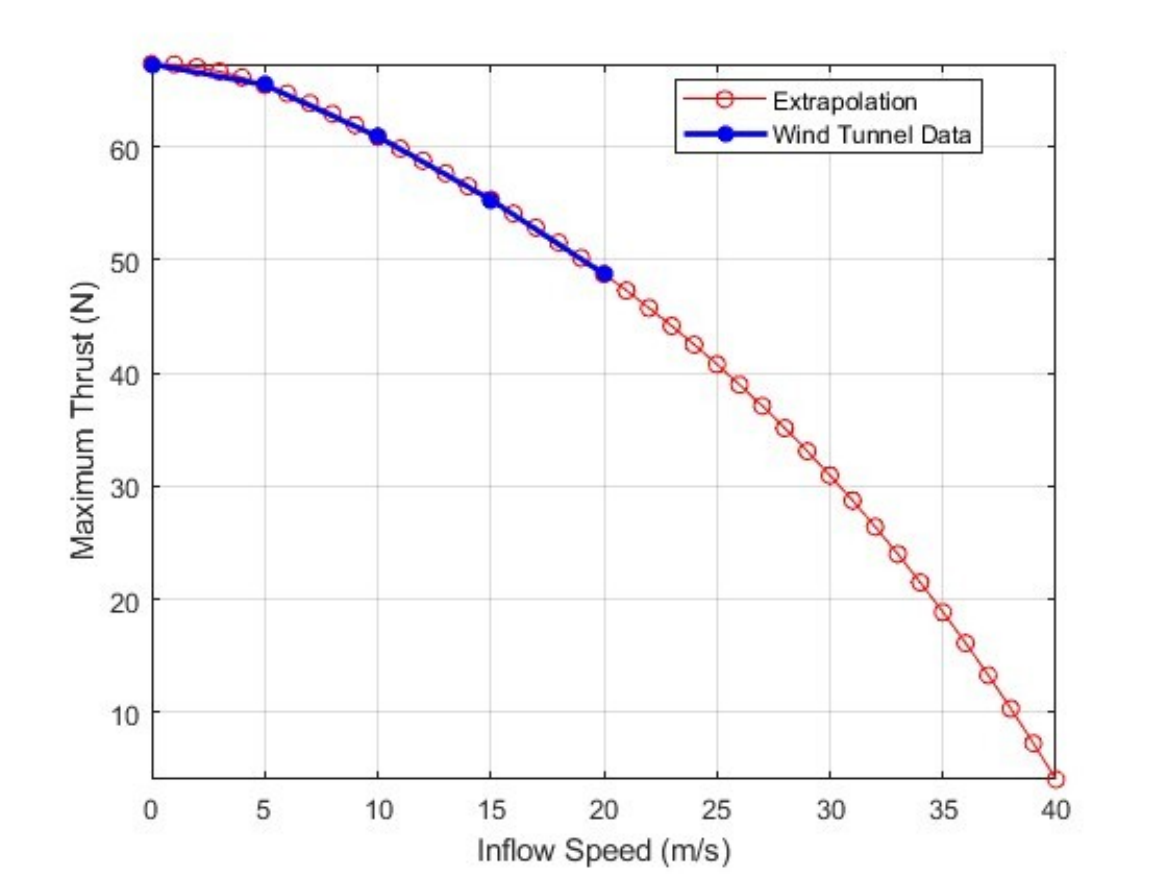}
        \caption{Thrust Maximum Performance}
        \label{fig_VDT_Wind_Tunnel_Test_Maximum_Thrust_Plot}
    \end{subfigure}%
    \caption{Wind Tunnel Facilities}
    \label{fig_VDT_Wind_Tunnel_Test}
    \vspace{-10pt}
\end{figure*}

\begin{table}[!htbp]
    \footnotesize
    \centering
    \caption{Wind Tunnel Test data}
    \label{tbl_VDT_Wind_Tunnel_Test_Data}
    \begin{tabular}{lccccc}
        \hline
        Inflow Speed ($m/s$) & 0 & 5 & 10 & 15 & 20  \\
        \hline
        Maximum Thrust ($N$) & 67.3 & 65.5 & 60.9 & 55.3 & 48.8 \\ 
        \hline
    \end{tabular}
\end{table}

\begin{table}
    \footnotesize
    \centering
    \caption{Domains and analysis methods for aerodynamic coefficients}
    \label{tbl_Domain_Analysis_Method_for_Aerodynamic_Coefficients}
    \begin{tabular}{cccc}
        \hline \multirow{2}{*}{\begin{tabular}{c} 
                Analysis \\
                Tool
        \end{tabular}} & \multicolumn{2}{c}{ Variables } & \multirow{2}{*}{\begin{tabular}{c} 
                Number of \\
                Samples
        \end{tabular}} \\
        & $\alpha$ & $\beta$ & \\
        \hline HETLAS & {$[-20,30]$} & {$[0,20]$} & 1200 \\
        \hline AVL & {$[-20,30]$} & {$[0,20]$} & 1200 \\
        \hline XFLR5 & {$[-20,30]$} & {$[0,20]$} & 1200 \\
        \hline Fluent & {$[-20,30]$} & {$[0,20]$} & 25 \\
        \hline
    \end{tabular}
\end{table}

\begin{figure*}[htbp]
    \footnotesize
    \centering
    \begin{subfigure}[b]{0.455\textwidth}
        \centering
        \includegraphics[width=0.85\textwidth]{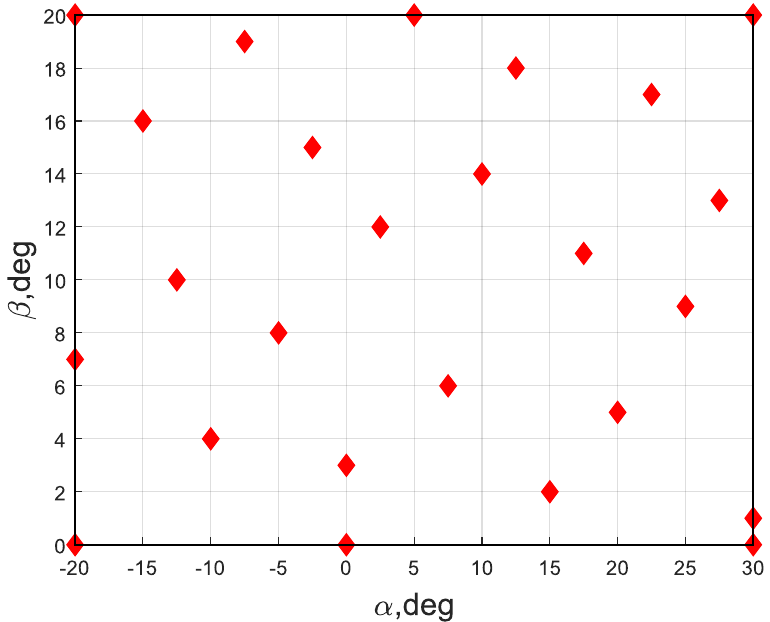}
        \caption{}
        \label{fig_VDT_Design_of_Experiments}
    \end{subfigure}
    \hfill
    \begin{subfigure}[b]{0.455\textwidth}
        \centering
        \includegraphics[width=0.7\textwidth]{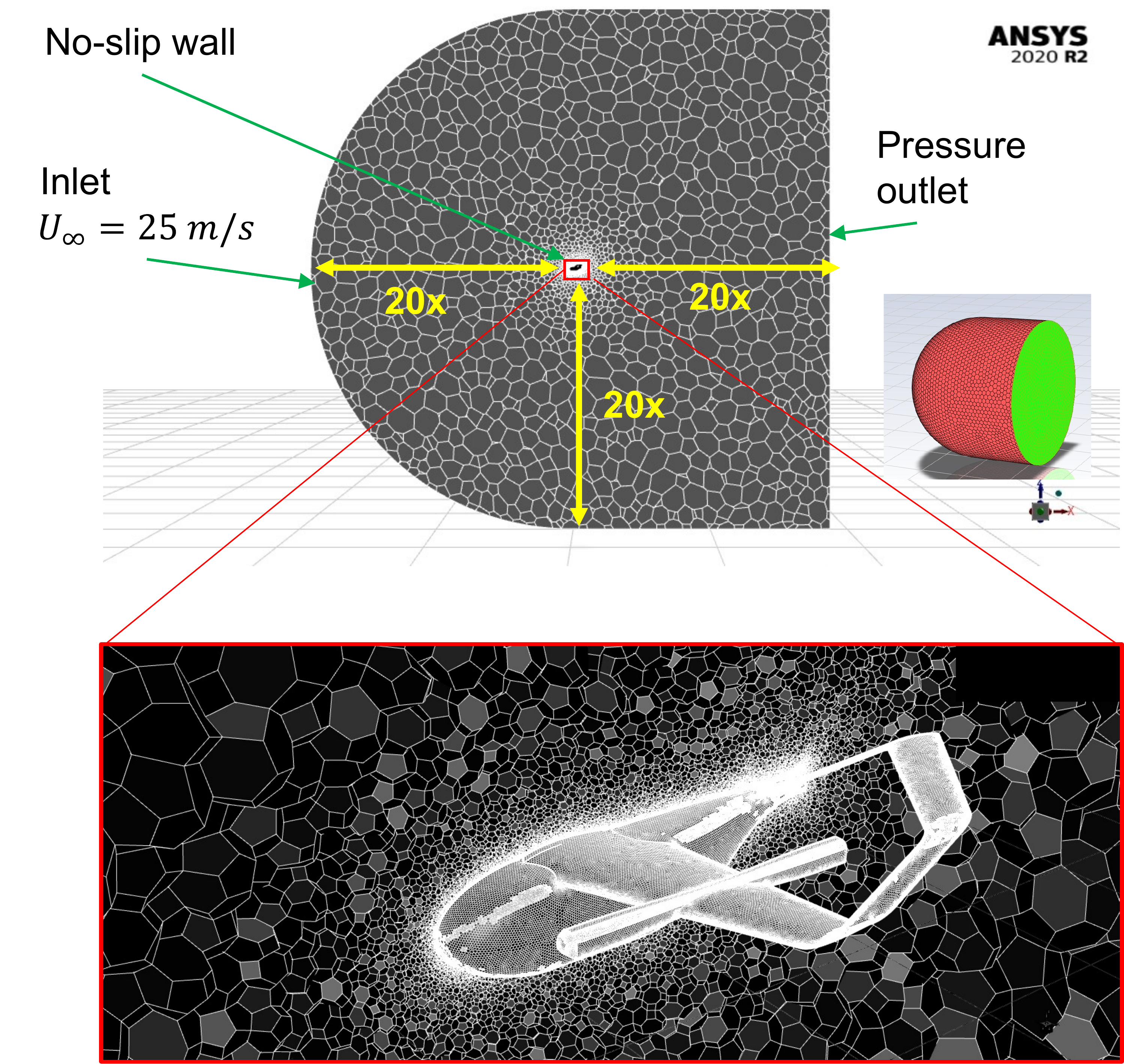}
        \caption{}
        \label{fig_VDT_a_Ansys_Boundary}
    \end{subfigure}
    \hfill
    \begin{subfigure}[b]{0.5\textwidth}
        \centering
        \includegraphics[width=\textwidth]{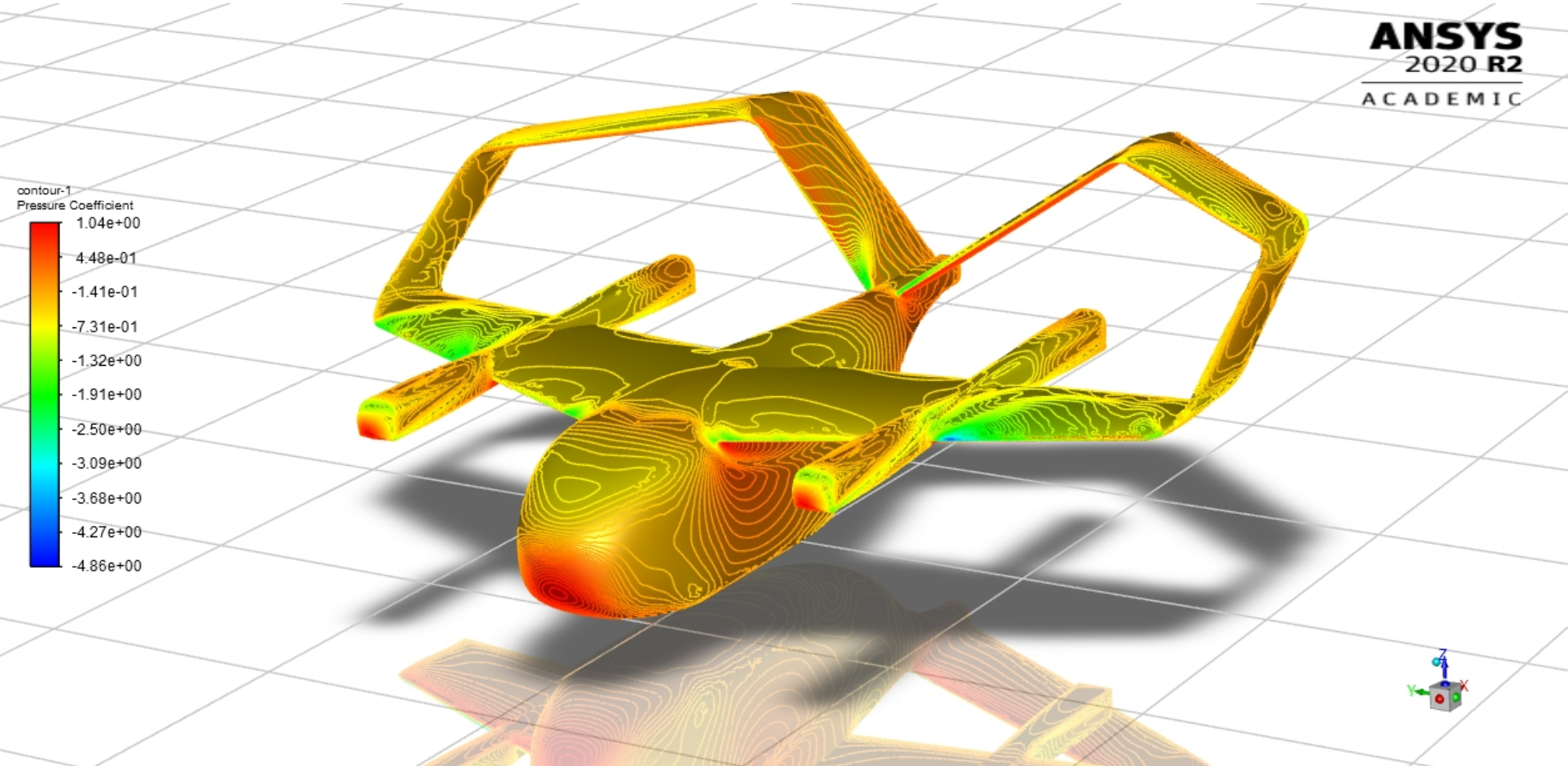}
        \caption{}
        \label{fig_VDT_b_Ansys_Pressure_Distribution}
    \end{subfigure}
    \caption{\footnotesize DOE and CFD Analyses; (a) Design of experiment for CFD analysis (b) Presentation of boundary conditions, far-field and prism boundary layer mesh of fuselage and wing. (c) Visualization of simulation results depicting pressure distribution on the surface.}
    \label{fig_VDT_CFD_Analyses}
\end{figure*}

\begin{figure}[!htb]
    \centering
    \includegraphics[width=1.0\textwidth]{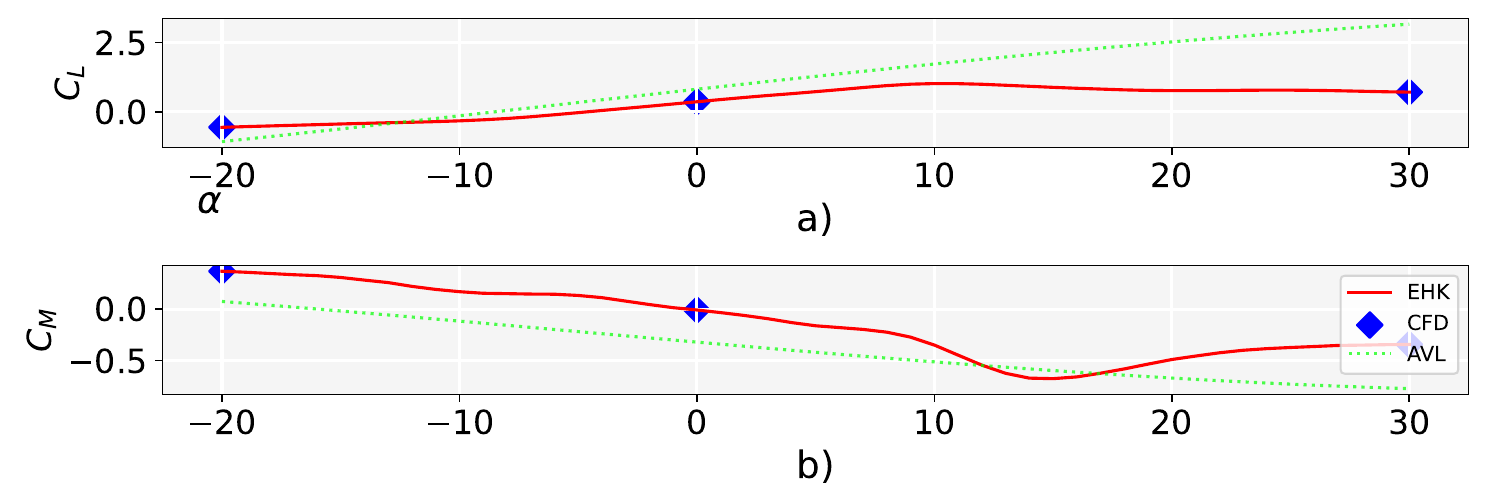}
    \caption{Lift and pitching moment coefficients due to $\alpha$ regarding various datasets }
    \label{fig_VDT_CM_CL_AOA}
\end{figure}

\begin{figure}[!htb]
    \centering
    \includegraphics[width=1.0\textwidth]{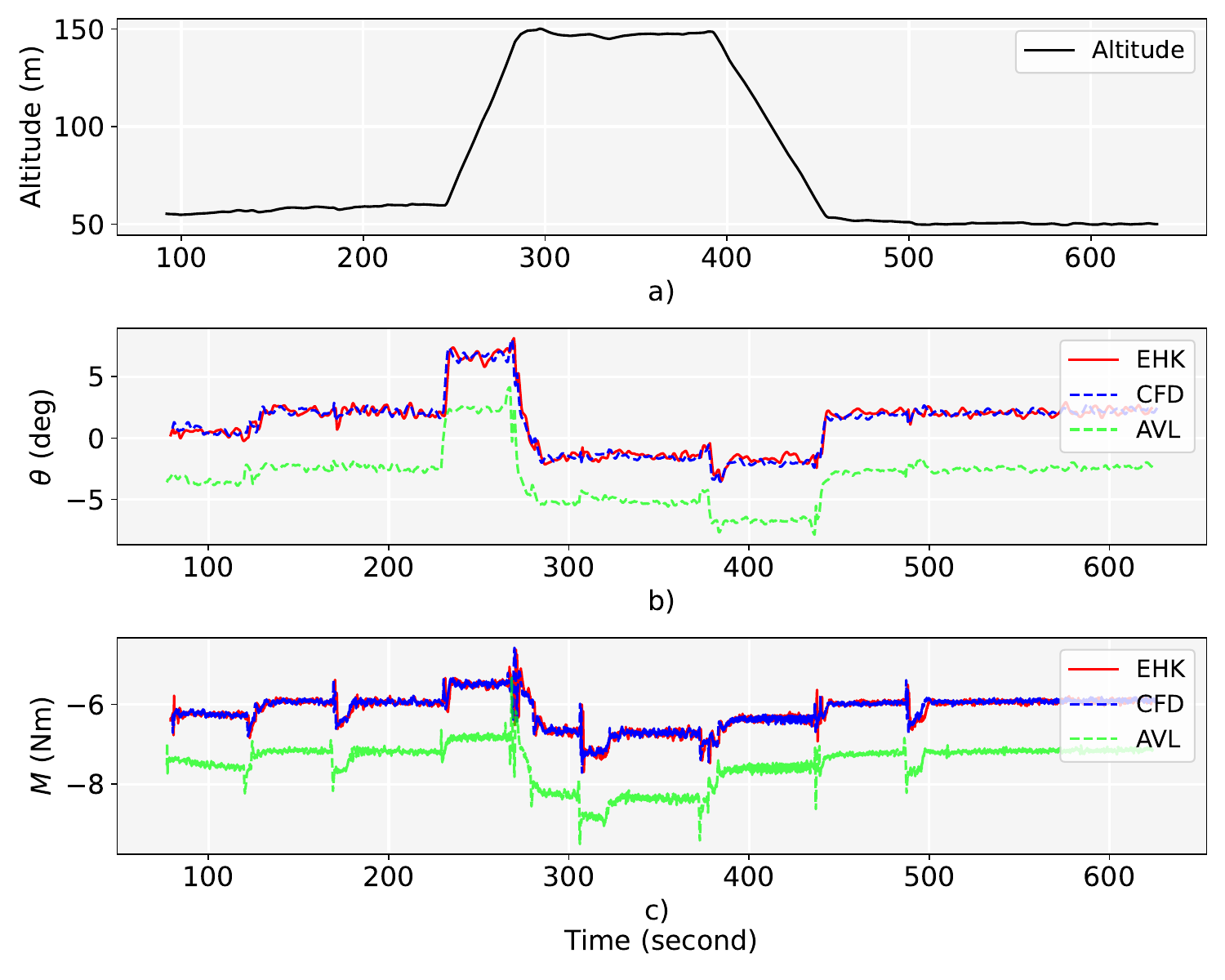}
    \caption{Altitude, pitching angle and moment of simulation model in time series}
    \label{fig_VDT_pitch_stability_analysis}
    \vspace{-10pt}
\end{figure}

\begin{figure*}[htbp]
    \centering
    \begin{subfigure}[b]{0.455\textwidth}
        \centering
        \includegraphics[width=1.0\textwidth]{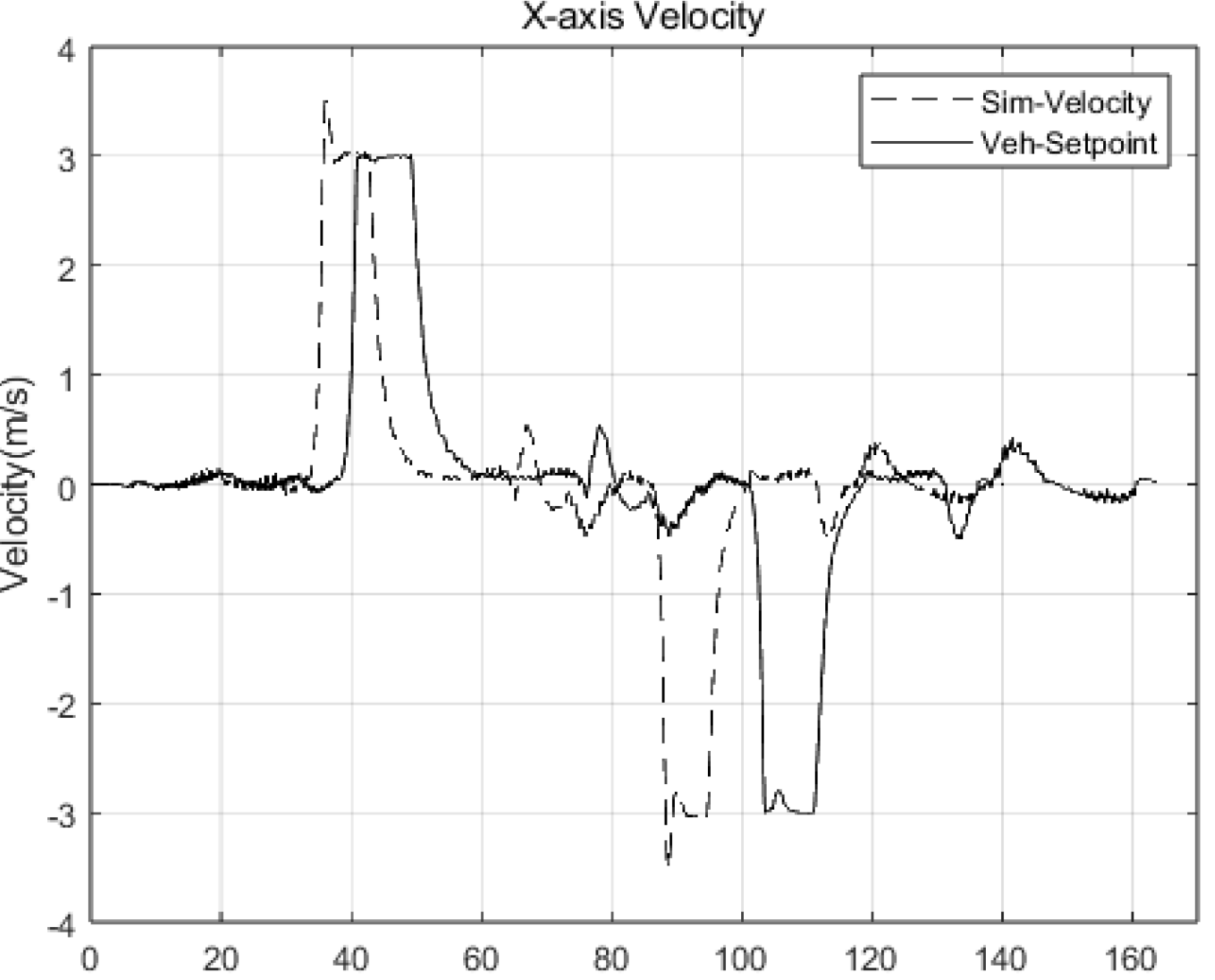}
        \caption{X-axis velocity}
        \label{fig_VDT_Results_X_Velocity_Digital_Layer_Setpoint_Physical_Layer}
    \end{subfigure}
    \hfill 
    \begin{subfigure}[b]{0.455\textwidth}
        \centering
        \includegraphics[width=1.0\textwidth]{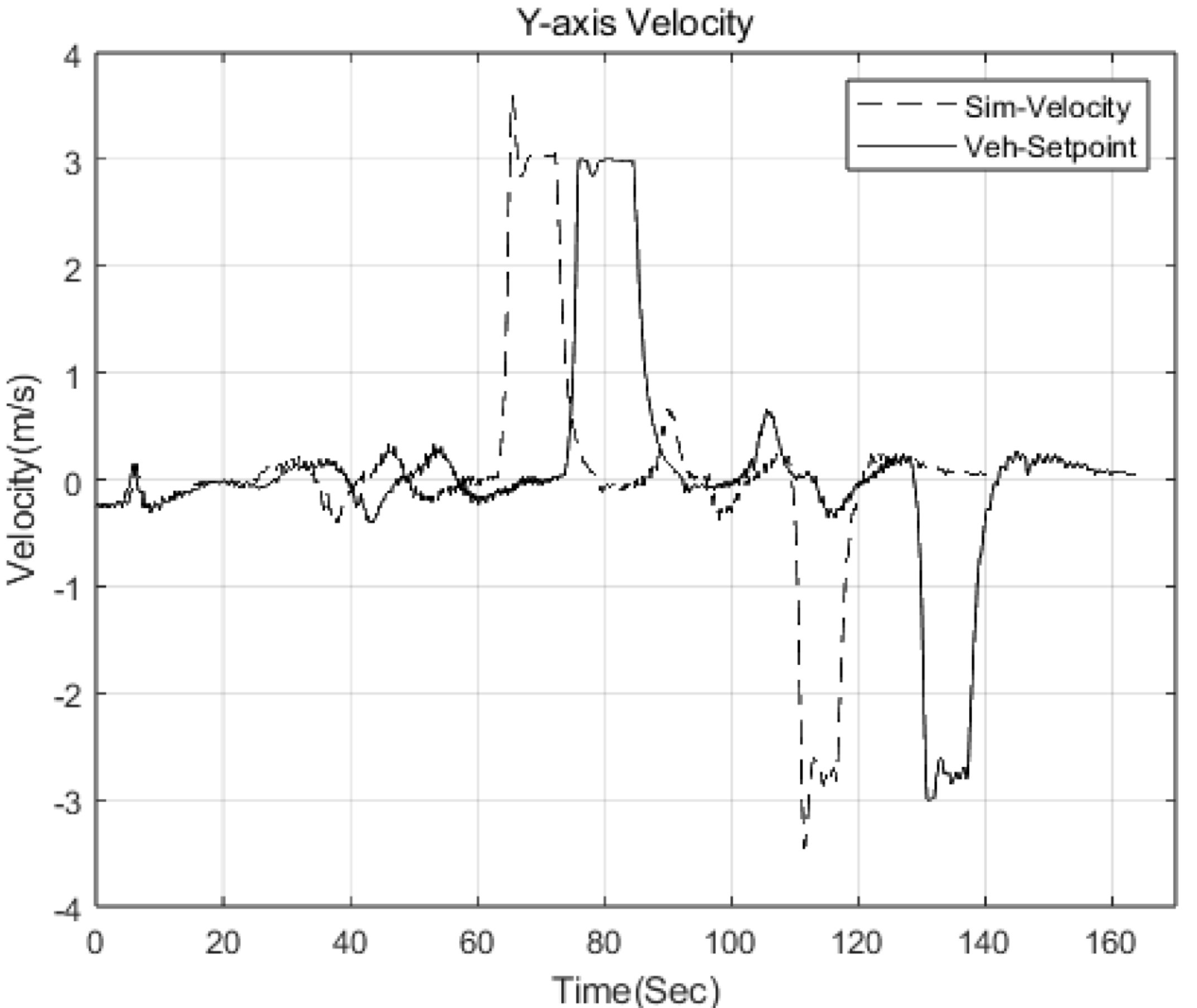}
        \caption{Y-axis velocity}
        \label{fig_VDT_Results_Y_Velocity_Digital_Layer_Setpoint_Physical_Layer}
    \end{subfigure}
    \hfill 
    \begin{subfigure}[b]{0.455\textwidth}
        \centering
        \includegraphics[width=1.0\textwidth]{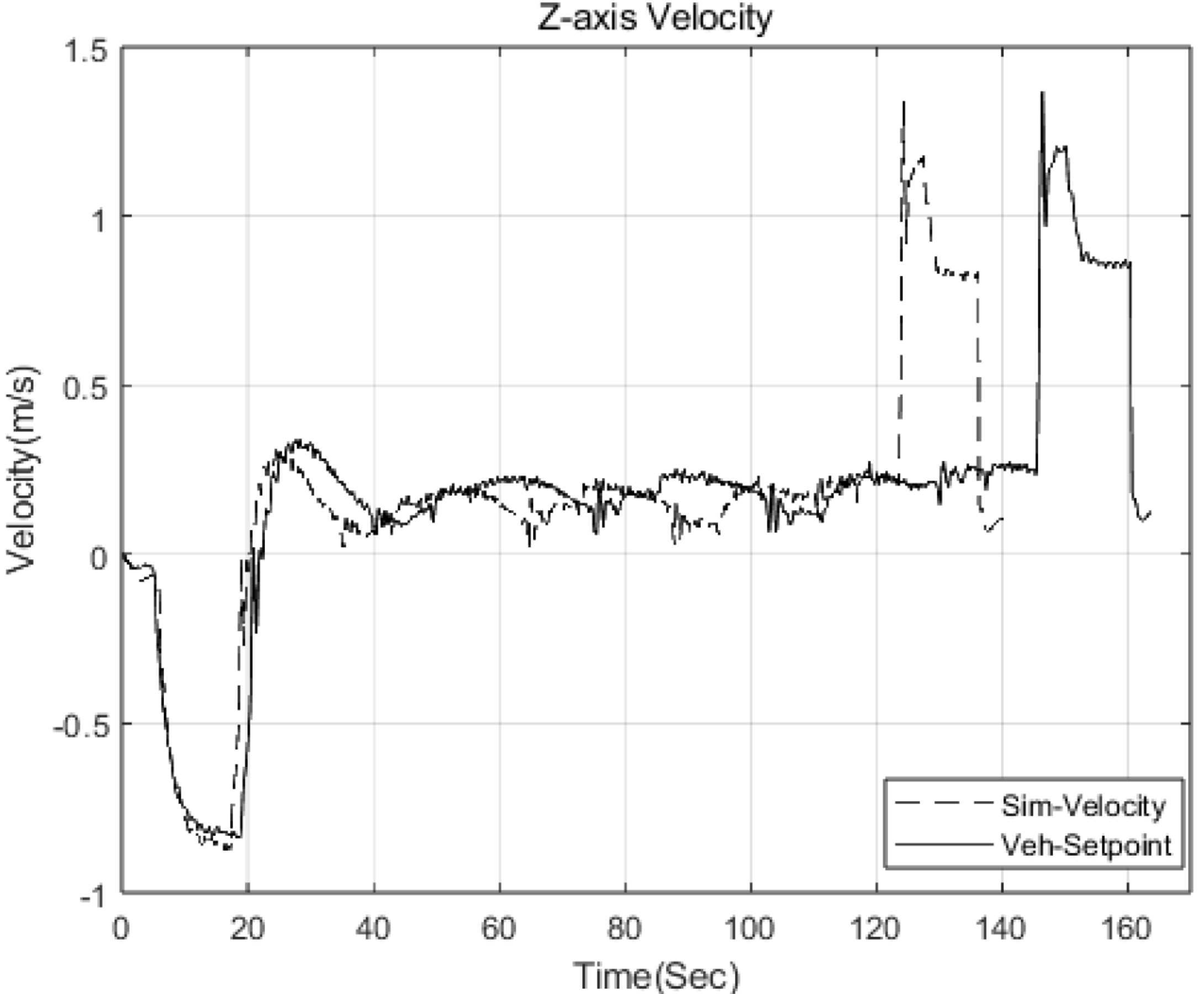}
        \caption{Z-axis velocity}
        \label{fig_VDT_Results_Z_Velocity_Digital_Layer_Setpoint_Physical_Layer}
    \end{subfigure}
    \hfill 
    \begin{subfigure}[b]{0.455\textwidth}
        \centering
        \includegraphics[width=1.0\textwidth]{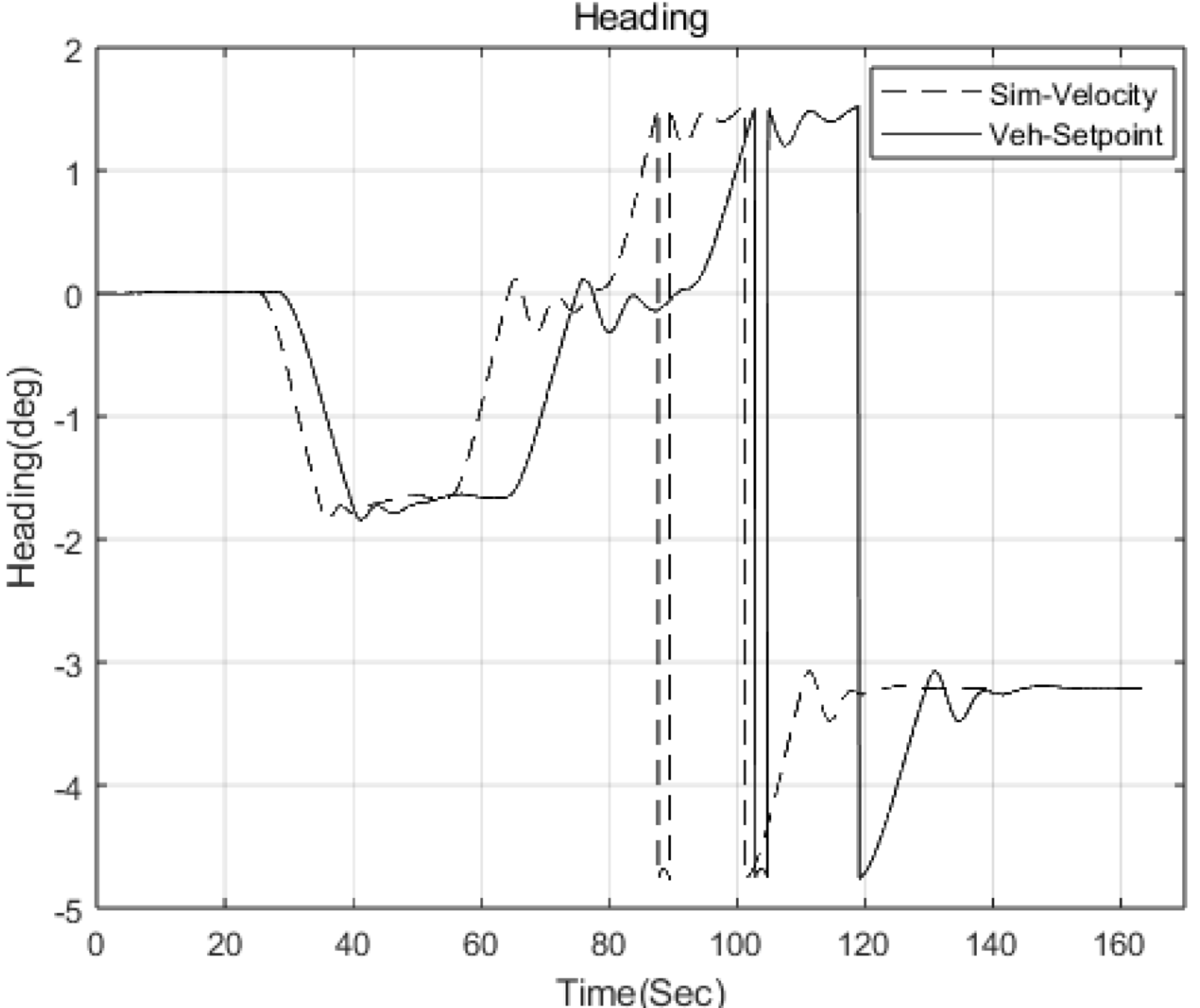}
        \caption{Heading}
        \label{fig_VDT_Results_Heading_Digital_Layer_Setpoint_Physical_Layer}
    \end{subfigure}
    \caption{Perform Velocity Commands to Digital Layer and Setpoints to Physical Layer}
    \label{fig_VDT_Results_Sending_Velocity_Digital_Layer_Setpoint_Physical_Layer}
    \vspace{-10pt}
\end{figure*}

\begin{figure*}[htbp]
    \footnotesize
    \centering
    \begin{subfigure}[b]{0.425\textwidth}
        \centering
        \includegraphics[width=1.0\textwidth]{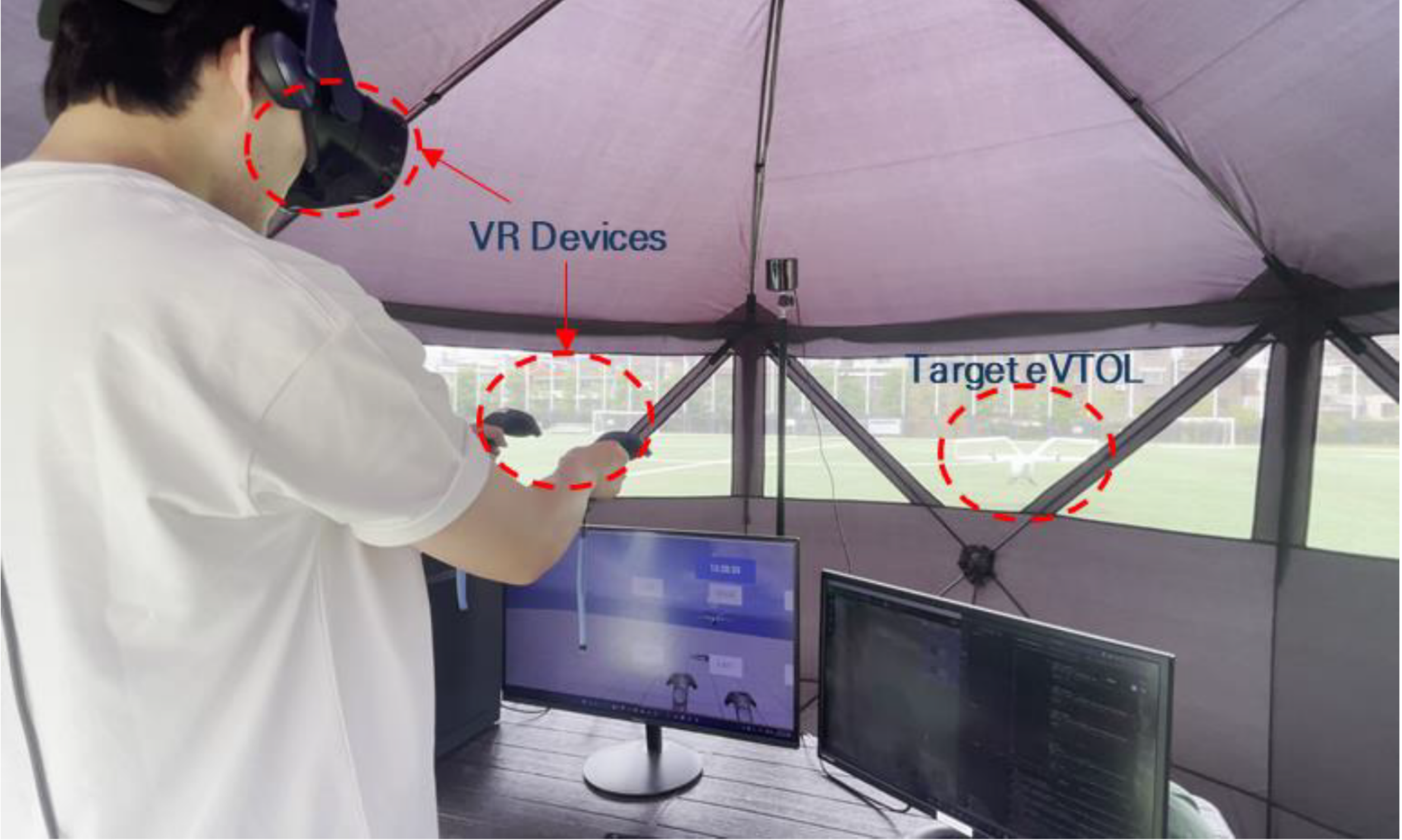}
        \caption{Actual VDT Flight Test}
        \label{fig_VDT_Flight_Test_Environment_Setup}
    \end{subfigure}
    \hfill 
    \begin{subfigure}[b]{0.555\textwidth}
        \centering
        \includegraphics[width=1.0\textwidth]{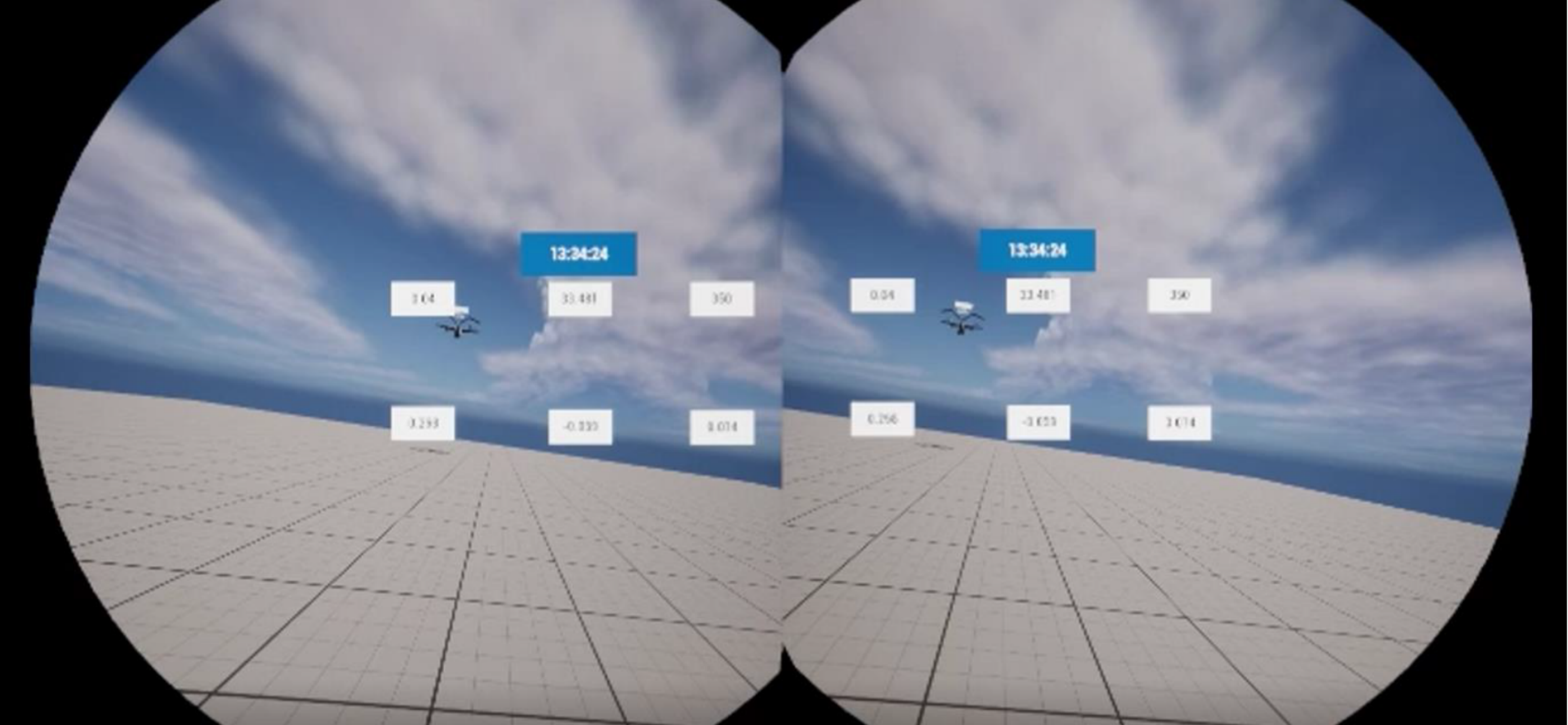}
        \caption{Binocular view of HMD}
        \label{fig_VDT_Results_VR_HMD_View_Flight_Test}
    \end{subfigure}
    \hfill 
    \begin{subfigure}[b]{\textwidth}
        \centering
        \includegraphics[width=1.0\textwidth]{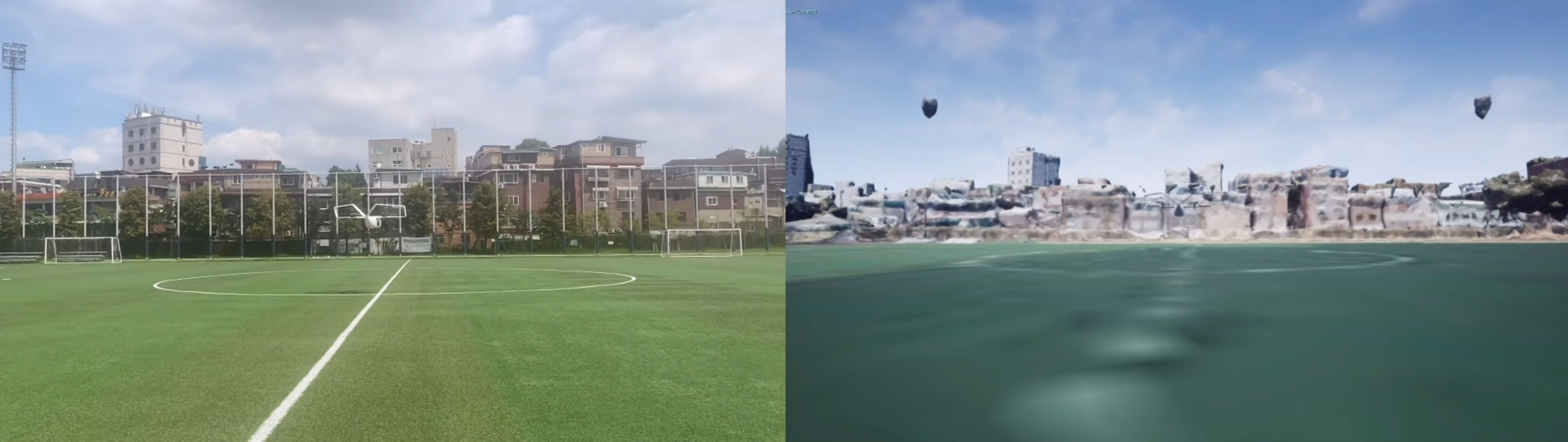}
        \caption{D2P Synchronization}
        \label{fig_VDT_D2P_Demo_Time_Step_01}
    \end{subfigure}
    \caption{Actual Flight Test }
    \label{fig_VDT_results_flight_test}
    \vspace{-10pt}
\end{figure*}

\section{Experiments and Results}
\label{sec_experiments_results}

\subsection{Experimental Propulsion Data}

The vehicle's electric propulsion system incorporates KDE Direct's \texttt{KDE5215XF-435} Brushless Direct Current (BLDC) motors and XOAR's PJN propellers, featuring a \texttt{15-inch} diameter and \texttt{7-inch} pitch. For the KP-2, carbon VTOL propellers and \texttt{5215XF} motors were selected. The dual battery configuration, each being \texttt{4S 16000 mah}, effectively doubles the capacity when connected in parallel. Wind tunnel experiments at Konkuk University, progressing in \texttt{5$m/s$} increments up to \texttt{20$m/s$}, facilitated the propulsion system's performance evaluation, as detailed in Table \ref{tbl_VDT_Wind_Tunnel_Test_Data}. These tests aimed at pinpointing peak performance across velocities, with high-speed scenarios estimated via cubic spline interpolation due to testing constraints. Thrust measurements and photographic evidence from these tests are showcased in Figures \ref{fig_VDT_Wind_Tunnel_Test}, underscoring the methodological approach to assessing single propulsion system efficacy within controlled wind tunnel environments.

\subsection{Simulation database fidelity analysis and verification}
In this investigation, aerodynamic analyses were executed for two key variables, the angle of attack ($\alpha$) and side-slip angle ($\beta$), confined within $-20^{\circ} \leq \alpha \leq 30^{\circ}$ and $0^{\circ} \leq \beta \leq 20^{\circ}$ respectively, at a constant velocity of $25 \mathrm{~m/s}$. Initial data were procured using Latin hypercube sampling (LHS) across high-fidelity (HF) and low-fidelity (LF) spectrums, with HF data derived from 25 CFD simulations via ANSYS-Fluent 2020 R2, and LF data aggregated from HETLAS, AVL, and XFLR5 tools, cumulatively yielding 3600 LF samples. These tools, despite their lower fidelity, afford broad condition coverage at reduced computational costs.

Figures \ref{fig_VDT_a_Ansys_Boundary} and \ref{fig_VDT_b_Ansys_Pressure_Distribution} illustrate the KP-2 aircraft’s 1/4-scale simulation model and its $10,970,498$-cell unstructured CFD mesh, encapsulating the aircraft within a domain scaled twentyfold relative to its fuselage in all dimensions. Simulation setups aligned the aircraft per respective $\alpha$ and $\beta$ angles, with a semi-spherical domain face and lateral cylinder wall imposing constant velocity and pressure outlet conditions applied domain-end. Figure \ref{fig_VDT_CM_CL_AOA} contrasts databases from diverse methodologies, revealing the CFD's limited condition span, AVL's expansive but low-accuracy data, and the EHK's integration of all datasets to furnish a comprehensive, high-accuracy aerodynamic coefficient database, as detailed in Table \ref{tbl_Domain_Analysis_Method_for_Aerodynamic_Coefficients}. Time series response analyses from Figure \ref{fig_VDT_pitch_stability_analysis} delineate notable response discrepancies among databases within identical mission profiles. Specifically, AVL predicted excessive lift and nose-down moments, manifesting in unrealistic negative pitching angles during takeoff-climbing and landing phases, diverging markedly from the more plausible positive pitching angles presented by EHK and CFD models. The AVL model exhibited pronounced negative pitching angles during cruising and descending phases, contrasting the minimal variations of EHK and CFD models, corroborated by pitching moment observations, which highlighted the AVL model's exaggerated nose-down tendencies. 

These findings accentuate the Extended Hierarchical Kriging (EHK) method's superiority in amalgamating disparate data sources to create a robust AeroDB, outperforming individual datasets in quality and coverage. The EHK and CFD models, underpinned by CFD data, displayed more consistent aerodynamic behavior, underscoring the EHK method's efficacy in enhancing simulation accuracy by bridging data gaps inherent to singular data sources. This approach validates the EHK method’s utility in synthesizing high-quality AeroDBs, emphasizing its potential to significantly improve aerodynamic simulation fidelity across varied flight conditions.

\subsection{Tele-operation Experiments}

Before implementing the digital twin's immersive teleoperation system on an actual eVTOL aircraft, a pre-validation was carried out within a simulation framework to evaluate the system's capability for synchronized flight control via the D2P process. This precautionary step aimed to identify and mitigate potential risks, ensuring system readiness for live flight trials. The integration with a SITL simulation, executed on distinct computing units linked by the User Datagram Protocol (UDP) and facilitated through MAVLinkRouter, mirrored the teleoperation system’s digital layer with the eVTOL's physical counterpart. The critical $POSITION\_TARGET\_LOCAL\_NED$ message was transmitted at a frequency of $30 \mathrm{~Hz}$, reflecting expected real-life data link performance.

The validation showcased the system’s precision in executing real-time command transfers from the digital domain to the physical eVTOL, following a predetermined square flight pattern. This underlined the system's adeptness in managing accurate speed commands across various axes, confirming synchronization capabilities despite observed time lags attributed to hardware constraints. Adjustments were also necessary to align with the eVTOL firmware's speed limitations, excluding velocities beyond $3 \mathrm{~m/s}$.

Employing a Virtual Reality Head-Mounted Display (VR HMD), the operator navigated the eVTOL’s flight dynamics, issuing control commands that were accurately mirrored by the physical aircraft, albeit with notable telemetry system-induced delays. This process validated the seamless interaction between the digital and physical realms, reinforcing the teleoperation system's functionality and operational integrity ahead of actual flight testing.

\section{Conclusion}
This study developed a VDT system for eVTOL aircraft in AAM, significantly improving remote piloting capabilities. The integration of digital twin technology with immersive VR interfaces enhances situational awareness and control precision, especially for BVLOS operations. Utilizing a high-fidelity digital replica of the eVTOL within a realistic simulated environment enables precise monitoring and control by remote operators. The research validates the system's ability to accurately transmit control commands and synchronize the digital and physical states of the eVTOL, demonstrating its potential to enhance operational efficiency and safety in AAM.

\section*{Acknowledgments}
\AckUAM

\bibliography{refs}

\begin{thebibliography}{5}
\newcommand{\enquote}[1]{``#1''}
\providecommand{\natexlab}[1]{#1}
\providecommand{\url}[1]{\texttt{#1}}
\providecommand{\urlprefix}{URL }
\expandafter\ifx\csname urlstyle\endcsname\relax
  \providecommand{\doi}[1]{\discretionary{}{}{}https://doi.org/#1}\else
  \providecommand{\doi}[1]{\discretionary{}{}{}\urlstyle{rm}\url{https://doi.org/#1}}\fi

\bibitem[{Licata and Cole(2024)}]{Licata2024}
Licata, R., and Cole, T., \enquote{{Pioneering Advanced Air Mobility},}
  \emph{AIAA SCITECH 2024 Forum}, American Institute of Aeronautics and
  Astronautics, Reston, Virginia, 2024.
\newblock \doi{10.2514/6.2024-1527},
  \urlprefix\url{https://arc.aiaa.org/doi/10.2514/6.2024-1527}.

\bibitem[{Ywet et~al.(2024)Ywet, Maw, Nguyen, and Lee}]{Ywet2024}
Ywet, N.~L., Maw, A.~A., Nguyen, T.~A., and Lee, J.-W.,
  \enquote{{YOLOTransfer-DT: An Operational Digital Twin Framework with Deep
  and Transfer Learning for Collision Detection and Situation Awareness in
  Urban Aerial Mobility},} \emph{Aerospace}, Vol.~11, No.~3, 2024, p. 179.
\newblock \doi{10.3390/aerospace11030179},
  \urlprefix\url{https://www.mdpi.com/2226-4310/11/3/179}.

\bibitem[{Jang et~al.(2023)Jang, Hyun, Kwag, Gwak, Nguyen, and Lee}]{Jang2023}
Jang, M., Hyun, J., Kwag, T., Gwak, C., Nguyen, T.~A., and Lee, J.-W.,
  \enquote{{Robust Attitude Control for PAVs using DNN with Exponentially
  Stabilizing Control Lyapunov Functions},} \emph{AIAA SCITECH 2023 Forum},
  American Institute of Aeronautics and Astronautics, Reston, Virginia, 2023.
\newblock \doi{10.2514/6.2023-1443}.

\bibitem[{Pham et~al.(2023{\natexlab{a}})Pham, Tyan, Nguyen, Lee, Nguyen, and
  Lee}]{PHAM2023}
Pham, V., Tyan, M., Nguyen, T.~A., Lee, C.-H., Nguyen, L.~T., and Lee, J.-W.,
  \enquote{{Adaptive data fusion framework for modeling of non-uniform
  aerodynamic data},} \emph{Chinese Journal of Aeronautics}, Vol.~36, No.~7,
  2023{\natexlab{a}}, pp. 316--336.
\newblock \doi{10.1016/j.cja.2023.05.012},
  \urlprefix\url{https://linkinghub.elsevier.com/retrieve/pii/S1000936123001644}.

\bibitem[{Pham et~al.(2023{\natexlab{b}})Pham, Tyan, Nguyen, and
  Lee}]{Pham2023_02}
Pham, V., Tyan, M., Nguyen, T.~A., and Lee, J.-W., \enquote{{Extended
  Hierarchical Kriging Method for Aerodynamic Model Generation Incorporating
  Multiple Low-Fidelity Datasets},} \emph{Aerospace}, Vol.~11, No.~1,
  2023{\natexlab{b}}, p.~6.
\newblock \doi{10.3390/aerospace11010006},
  \urlprefix\url{https://www.mdpi.com/2226-4310/11/1/6}.

\end{thebibliography}

\end{document}